\documentclass[12pt]{article}
\usepackage{scicite}
\usepackage{times}
\usepackage{graphicx}
\graphicspath{{Figures/}}
\usepackage{setspace}

\usepackage{hyperref}
\usepackage{cleveref}
\crefname{equation}{Eq.}{Eqs.}
\Crefname{equation}{Equation}{Equations}
\crefname{figure}{Fig.}{Figs.}
\Crefname{figure}{Figure}{Figures}
\crefname{section}{Sect.}{Sects.}
\Crefname{section}{Section}{Sections}
\crefname{table}{Table}{Tables}
\crefname{appsec}{Appendix}{Appendices}

\newenvironment{sciabstract}{
\begin{quote} \bf}
{\end{quote}}

\title{\large Experimental realization of an intrinsically error-protected \\ superconducting qubit} 

\author
{\normalsize Andr\'as Gyenis,$^{1\ast}$ Pranav S. Mundada,$^{1\ast}$ Agustin Di Paolo,$^{2\ast}$ Thomas M. Hazard,$^{1}$ 
\\
\normalsize Xinyuan You,$^{3}$ David I. Schuster,$^{4}$
Jens Koch,$^{5}$ Alexandre Blais,$^{2,6}$ Andrew A. Houck,$^{1\dagger}$\\
\\
\footnotesize{$^{1}$Department of Electrical Engineering, Princeton University, Princeton, New Jersey 08544, USA}\\
\footnotesize{$^{2}$Institut quantique \& D\'epartement de Physique, Universit\'e de Sherbrooke, Sherbrooke J1K2R1 Quebec, Canada}\\
\footnotesize{$^{3}$ Graduate Program in Applied Physics, Northwestern University, Evanston, Illinois 60208, USA}\\
\footnotesize{$^4$ The James Franck Institute and Department of Physics, University of Chicago, Chicago, Illinois 60637, USA}\\
\footnotesize{$^{5}$ Department of Physics and Astronomy, Northwestern University, Evanston, Illinois 60208, USA}\\
\footnotesize{$^{6}$Canadian Institute for Advanced Research, Toronto, M5G1M1 Ontario, Canada}\\ 
\\
\footnotesize{$^\ast$These authors contributed equally to this work.}\\
\footnotesize{$^\dagger$To whom correspondence should be addressed; E-mail:  aahouck@princeton.edu.}
}

\date{} 

\begin{document} 

\baselineskip22pt
\maketitle 

\begin{sciabstract}
  Encoding a qubit in logical quantum states with wavefunctions characterized by disjoint support and robust energies can offer simultaneous protection against relaxation and pure dephasing. Using a circuit-quantum-electrodynamics architecture, we experimentally realize a superconducting $\mathbf{0-\pi}$ qubit, which hosts protected states suitable for quantum-information processing. Multi-tone spectroscopy measurements reveal the energy level structure of the system, which can be precisely described by a simple two-mode Hamiltonian. We find that the parity symmetry of the qubit results in charge-insensitive levels connecting the protected states, allowing for logical operations. The measured relaxation (1.6 ms) and dephasing times (25 $\mu$s) demonstrate that our implementation of the $\mathbf{0-\pi}$ circuit not only broadens the family of superconducting qubits, but also represents a promising candidate for the building block of a fault-tolerant quantum processor.
\end{sciabstract}

Preserving the coherence of a quantum superposition over prolonged times is key for large-scale quantum-information processing \cite{nielsen,devoret,knill}.  For example, quantum error correction protects states by using a large number of physical qubits to encode a single logical qubit \cite{shor,reed, kelly, ofek}. A complementary approach to maintain coherence over long timescales is to develop qubits which are intrinsically protected against decoherence \cite{ioffe,doucot,kitaev2,kitaev1,brooks,dempster,groszkowski,dipaolo,doucot2,gladchenko,bell,smith}. Such protection arises because quantum information in these circuits is encoded in delocalized collective states capable of withstanding errors originating from local noise. One of the most promising candidates for a fully-protected qubit is the $0-\pi$ circuit (Fig.~1A) proposed by Kitaev, Brooks and Preskill \cite{kitaev1,brooks}. However, protected circuits generally impose taxing requirements on the parameters of the physical device that are beyond the feasibility of current technologies. In this work, we realize a slightly modified version of the $0-\pi$ qubit by reducing the energy scales of its parameters to an experimentally obtainable regime. We demonstrate exponential protection against relaxation for the offset-charge-insensitive logical states, and show that dephasing due to flux noise is first-order suppressed. This manifests in significantly enhanced relaxation and coherence times and makes the $0-\pi$ circuit a leading contender for a superconducting quantum computer.  

Because the coherence of a qubit is affected by energy relaxation and pure-dephasing processes, protected qubits must be robust against both of these mechanisms. According to Fermi's golden rule, the decay rate of an excited state is proportional to the square of the matrix element that connects it to other states via a noise operator. Thus, a circuit with a Hamiltonian that has eigenstates with disjoint support can prevent rapid loss of information \cite{earnest,lin}. Such protection against relaxation, however, is insufficient to stabilize the phase of a superposition of logical states. The key idea of our work is to engineer a superconducting circuit where the energies of disjoint states are also robust against environmental noises. As we show here, this can be achieved in certain parameter regimes of the $0-\pi$ circuit. 

The $0-\pi$ qubit consists of identical pairs of small Josephson junctions, large shunting capacitors and superinductors, which are all organized in a single closed loop geometry with four nodes (Fig.~1A). The circuit has four degrees of freedom \cite{dempster}, which we refer to as the $\phi$, $\theta$, $\zeta$ and $\Sigma$ modes, and correspond to the linear combinations of the phase difference of the superconducting order parameter across the various elements in the circuit (Fig.~1C). Among them, the $\Sigma$ mode is cyclic, while the $\zeta$ mode represents a harmonic mode that decouples from the other modes in the absence of circuit-element disorder. The remaining $\phi$ and $\theta$ modes describe the qubit degrees of freedom of the circuit with the following two-mode Hamiltonian 
\begin{equation}
    H_{0-\pi}=4E_C^\theta(n_\theta - n_g^\theta)^2 +4E_C^\phi n_\phi^2 - 2E_J\cos\theta\cos(\phi-\pi{\Phi_\mathrm{ext}}/{\Phi_0}) + E_L\phi^2.
\end{equation}
Here,  $E^{\theta(\phi)}_C=e^2/2C_{\theta(\phi)}$  denotes the charging energy corresponding to the $\theta$ ($\phi$) mode with total capacitance of $C_{\theta(\phi)}$, $E_J$ is the Josephson energy, $E_L=\Phi_0^2/4\pi^2L$ is the inductive energy of the superinductor with inductance $L$, $\Phi_0=h/2e$ is the magnetic flux quantum (with $e$ the electron charge and $h$ the Planck's constant), $\Phi_\mathrm{ext}$ is the external magnetic flux threaded through the loop of the device, $n_g^\theta$ is the offset-charge bias due to the electrostatic environment, whereas $n_\theta$ and $n_\phi$ are the canonical charge operators corresponding to the phase operators (in units of $2e$). In the phase representation of quantum electromagnetic circuits \cite{vool}, the capacitive energies of the device determine the kinetic energies of the modes, while the staggered double well potential of the $0-\pi$ qubit is realized by the inductors and Josephson junctions (Fig.~1B).

Remarkably, the three physical modes of the $0-\pi$ circuit are analogous to the three fundamental representatives of superconducting qubits: the transmon\cite{koch}, the fluxonium\cite{manucharyan} and the cavity\cite{krastanov}. Indeed, as Fig.~1C shows, the $\theta$ mode describes the superconducting phase difference across the large shunting capacitors and the Josephson junctions, leading to a transmon-like behavior \cite{koch}. On the other hand, the $\phi$ mode corresponds to the phase drop across the Josephson junctions and the superinductors, which features a fluxonium-type response \cite{manucharyan}. Finally, the $\zeta$ mode arises from the phase difference across the superinductances and the shunting capacitors, resulting in a low-energy harmonic mode. 

As the intrinsic protection of the $0-\pi$ qubit emerges from the interplay of its effective double-well potential and the anisotropic kinetic energy of the modes \cite{kitaev1,brooks,dempster,groszkowski,dipaolo}, engineering the proper energy scales in the $0-\pi$ qubit is crucial. We first consider the case of protection against energy relaxation, which is provided by localizing the qubit wavefunctions in either the $\theta=0$ or in the $\theta=\pi$ valley (Fig.~1B). The circuit realizes a double-well potential with two fluxonium-like potentials, $V(\phi,\theta=0,\pi)=\mp2E_J\cos(\phi)+E_L\phi^2$, along the $0$ and $\pi$ valleys. The two potentials are displaced with respect to each other such that the $\theta=0$ valley has a single minimum at $\phi=0$, whereas the $\theta=\pi$ valley features two minima around $\phi\simeq\pm\pi$. Importantly, the potential energy difference between the valleys ($E_L\pi^2$) due to the quadratic inductive term corresponds approximately to the transition frequency between the ground states of the two valleys, i.e., the logical qubit energy. To ensure that the logical excited state is localized along the $\theta$ direction, first, the effective barrier height separating the valleys ($\sim4 E_J$) is required to be  much larger than the qubit transition energy, which we realize with $E_J/E_L\approx 16$. Second, as the tunneling amplitude between the valleys is exponentially reduced with the ratio of barrier height and kinetic energy \cite{koch},  we choose $E_J/E_C^\theta\approx 65$.

As the $0-\pi$ qubit couples to charge and flux degrees of freedom through the $\theta$ and $\phi$ modes, we achieve protection from dephasing by taking advantage of the two-dimensional nature of the potential to combine the beneficial parameter regimes of the transmon and fluxonium qubits. First, to exponentially suppress the charge sensitivity of the qubit, we simply operate the compact $\theta$ mode in the transmon regime with $E_J/E_C^\theta\approx 65$\cite{koch}. Second, to overcome the flux sensitivity, we exploit the avoided crossing of the two lowest-lying levels of $\theta=\pi$ valley to engineer a first-order-insensitive magnetic sweet spot (Fig.~1D). Indeed, the presence of the $\phi\rightarrow-\phi$ symmetry at $\Phi_\mathrm{ext}=0$ accompanied by two local minima in the potential of the $\pi$ valley leads to degenerate doublets, which are hybridized due to the finite kinetic energy along the $\phi$ direction. Such hybridized (symmetric and antisymmetric) states show a hyperbolic dispersion as a function of external flux and a first-order-insensitive sweet spot at zero field (Fig.~1D). The gap size of the avoided crossing and consequently the protection against flux noise, is proportional to the tunneling rate between the two local minima of the $\pi$ valley,  therefore requiring a sufficiently large $E_C^\phi$. Ultimately, the $\theta$ and $\phi$ modes are rendered heavy and light, respectively, by an anisotropic kinetic-energy ratio of $E_C^\phi/E_C^\theta\approx12$. 

Figure 1E shows the energy spectrum and wavefunctions of our $0-\pi$ circuit in the absence of external magnetic fields. In the 0 valley, the excitations are plasmon-like with wavefunctions similar to those of an anisotropic, two-dimensional harmonic oscillator. The nodes of the wavefunctions appear first along the $\theta$ direction as the kinetic energy of the $\theta$ mode is much lower than that of the $\phi$ mode. In the $\pi$ valley, the states appear in symmetric-antisymmetric pairs, with nodes again developing first along the $\theta$ direction. Inspired by the quantum numbers of natural atoms, we denote low-lying energy levels as $|n_{lm}^j\rangle$, where the first quantum number $n=0,\pi$ refers to the valley index, $l=s,p,d,f\ldots$ specifies the number of the nodes of the wavefunction, and $m=\theta,\phi$ determines the orientation of the nodes. Finally, for the states in the $\pi$ valley, the superscript $j=+,-$, refers to the $\phi$-parity of the state. In this work, we use the ground states of the two  valleys, $|0_s\rangle$ and $|\pi_s^+\rangle$, as the logical qubit states.

We fabricated the $0-\pi$ device using conventional lithographic techniques in a two-dimensional circuit-QED architecture (Fig.~2A and B). To probe the qubit using dispersive readout\cite{wallraff,blais}, we capacitively coupled it to a coplanar-waveguide cavity with resonant frequency $\omega_c/2\pi=7.328$ GHz and photon decay rate $\kappa/2\pi=1.6$ MHz. As Fig.~2B shows, our primary goal for the circuit layout is to implement the highly anisotropic nature of the kinetic energies of the $\theta$ and $\phi$ modes with two tightly interdigitated niobium capacitors placed at a large distance from each other. Although this design increases the susceptibility of the device to dielectric losses due to the extremely small gap between the capacitor fingers (600 nm), it reduces the cross capacitances contributing to the light $\phi$ mode while maintaining a large enough capacitance ($C=101$ fF) for the heavy $\theta$ mode, yielding $E_C^\theta/h=92$ MHz and $E_C^\phi/h=1.14$ GHz.  The two small Josephson junctions are double-angle evaporated Al-AlO$_{x}$-Al Dolan-type junctions with $E_J/h = 6.0$ GHz, and each superinductor is realized by an array of 200 large Josephson junctions resulting in $E_L/h=0.38$ GHz. We choose a hybrid resonator-coupling scheme where all four nodes of the qubit have considerable coupling capacitance to both the centerpin of the resonator and the ground plane, which allows us to address both the $\phi$ and $\theta$ modes in our measurements \cite{supplement}. The value of these coupling capacitances are carefully chosen to realize sufficiently large coupling rates for the qubit operation while minimizing stray capacitances associated with the light $\phi$ mode. Additionally, DC voltage-biasing the centerpin of the resonator allows us to tune the offset charges on the islands of the device. 

To map out the energy spectrum of the $0-\pi$ qubit, we perform standard two-tone spectroscopy as a function of external magnetic flux and offset-charge bias. In the dispersive limit of circuit QED \cite{wallraff,blais}, we can probe the excitation of various transitions by monitoring the transmission at the cavity frequency while sweeping the frequency of a second spectroscopic tone. At low frequencies (Fig.~2C), we detect the response of the harmonic $\zeta$ mode \cite{masluk}, while at higher frequencies the excitations of the anharmonic qubit modes are probed (Fig.~2D to F). As expected, the spectroscopic data obtained as a function of flux (Fig.~2D and E) reveal two types of transitions: intra-valley plasmon transitions in the 0 valley, which are characterized by flat almost flux-independent dispersions, and inter-valley fluxon excitations between the 0 and $\pi$ wells, which have strong flux dependence. At offset charge $n_g^\theta = 0$ (Fig.~2D), we observe two distinct sets of transitions corresponding to the odd and even charge parity of the islands, which is the signature of the intermittent tunneling of unpaired quasiparticles across the junctions\cite{aumentado,schreier,sun,riste,serniak}. By contrast, at $n_g^\theta = 0.25$ (Fig.~2E), we observe only one set of transitions, which indicates the insensitivity of the qubit to individual quasiparticle tunneling events. The dependence of the transition frequency on the charge parity is more apparent when we measure the qubit spectrum as a function of $n_g^\theta$ (Fig.~2F). At low energies, the transmon-like excitations have exponentially suppressed charge dispersions\cite{koch,schreier}, where we are unable to resolve the different charge-parity states. At higher frequencies, however, “eye-like” patterns appear with dispersions up to $\sim$ 1 GHz due to the strong charge-sensitivity of the higher-lying levels. The spectral weight of the transitions with opposite parity is equal, which implies that both parities occur during the integration time of the spectroscopic measurements. 

It is worth emphasizing that we find remarkable agreement between the simple two-mode Hamiltonian in Eq.~(1) of the $0-\pi$ qubit and the experimental data over the entire range of both offset charge and external flux (solid dashed lines in Fig.~2D to F). The relatively simple theoretical model not only captures accurately the energy level structure with at least 17 transitions and over a 12 GHz frequency range, but also predicts the cavity-assisted sideband transitions and qubit transitions due to thermal occupation of low-lying levels \cite{supplement}. This excellent agreement highlights that although the $0-\pi$ artificial atom is constructed from the combination of $400$ Josephson junctions and large capacitors, its effective dynamics is fairly simple, which is an inevitable requirement for a qubit to be implemented in a large scale quantum processor.

Owing to the exponentially small dipole matrix element between the ground states of the valleys ($|0_s\rangle$ and $|\pi_s^+\rangle$), direct transitions between these protected states are strongly suppressed. To control the qubit, we therefore take advantage of higher energy states with support in both valleys. These levels, however, are more sensitive to offset charge: both the excitation energies and the transition dipole elements are dependent on $n_g^\theta$. 

We stress that higher-lying levels must have small charge dispersion and non-vanishing coupling to both logical states to serve as ancillary states for qubit operation. To shed light on how to simultaneously satisfy these two requirements, we adopt a simple band-structure picture based on the analogy between the periodic Coulomb potential of a solid crystal and the periodic potential of the compact $\theta$ mode of the $0-\pi$ circuit (Fig.~3). By extending the $0-\pi$ potential beyond the $\theta \in [0,2\pi)$ region, the $\theta$ phase can be understood as describing the position along a fictitious one-dimensional crystal in phase space (Fig.~3A), and the eigenfunctions are quasi-periodic Bloch states $\Psi_{n_g}(\theta,\phi)$  (Fig.~3B and C). In a tight-binding approximation, the charge dispersion takes the usual form $\Delta \epsilon (n_g^\theta) \approx 2t \cos(2\pi n_g^\theta)$ where $t$ is the hopping matrix element between localized atomic (Wannier) states. Similarly, drive-assisted transitions between qubit states can be expressed by transitions between neighboring Wannier states \cite{supplement}.

We first focus on the charge dispersion of the higher-lying levels and establish that states located mostly in the $\theta=\pi$ valley that are antisymmetric in $\phi$ are suitable intermediate levels for population transfer between the logical states. To show this symmetry-protected charge insensitivity, we carry out spectroscopic measurements as a function of $n_g^\theta$ on the members of the $|\pi_{d\theta}^\pm\rangle$ symmetric-antisymmetric states (Fig.~3D). We find that  while the symmetric state $|\pi_{d\theta}^+\rangle$ exhibits a strong charge dispersion, its antisymmetric partner $|\pi_{d\theta}^-\rangle$ is almost offset-charge insensitive. This behavior is in complete agreement with the tight-binding picture where the strongly localized $|\pi_{d\theta}^-\rangle$ state (Fig.~3E) results in a small hopping integral $t^-$ and a heavy flat band, in contrast to the light band associated with the more delocalzied $|\pi_{d\theta}^+\rangle$ state (Fig.~3F). We note that the different degrees of localization of the atomic states with opposite $\phi$ parity can be attributed to the anisotropic kinetic energies of the modes. This important observation is the foundation of our protocol for coherent control of the $0-\pi$ qubit where we use the charge-insensitive state $|\pi_{d\theta}^-\rangle$ as the ancillary level.

Unitary control relying on higher energy states imposes a second demand on the ancillary level: the transition matrix elements connecting the intermediate state to both logical states must be finite. Intriguingly, in the $0-\pi$ circuit these matrix elements have an anomalous offset-charge dependence with $4e$ periodicity, which is a manifestation of the Aharonov-Casher interference effect \cite{friedman,manucharyan2,pop,bell2,supplement}. To understand this feature of the inter-valley transitions, we again harness the tight-binding approximation. In this picture (Fig.~3G and H), there are two paths for the coherent drive to excite a fluxon transition: an initial Wannier state located in the 0 valley can be excited to a final Wannier state in the $\pi$ valley or in the $-\pi$ valley \cite{supplement}, while transitions to more distant valleys are strongly suppressed. The geometric phase \cite{wilczek} difference between the states in the $\pm\pi$ valleys leads to the offset-charge dependent interference pattern of the matrix element related to the double-Cooper-pair tunneling events. 

We experimentally measure the drive-assisted interference effect on the charge matrix element to find the optimal charge bias point of the $0-\pi$ qubit. In our scheme (Fig.~3I inset), we monitor a plasmon transition ($|0_{s}\rangle\rightarrow|0_{p\theta}\rangle$) with a weak probe tone, while irradiating the qubit with a strong coupler drive that addresses a fluxon transition ($|0_{p\theta}\rangle\rightarrow|\pi_{p\theta}^-\rangle$). The purpose of the probe tone is to map out the dressed states formed by driving the fluxon transition with the coupler tone. As Fig.~3I shows, when the coupler drive is on resonance with the fluxon frequency, the transition is split into two levels, known as the Autler-Townes doublet \cite{baur,sillanpaa,novikov}. The doublet is separated by the Rabi splitting $\Omega_c$, which is proportional to the voltage amplitude of the drive and the dipole matrix element of the transition. In fact, we observe a pair of Autler-Townes doublets for each offset-charge bias point \cite{supplement}, corresponding to the even or odd charge states due to the aforementioned quasiparticle poisoning. By keeping the drive strength constant and changing the induced bias $n_g^\theta$, we monitor the Rabi splitting of the fluxon state to determine the behavior of the charge matrix element.  This reveals an interference pattern in excellent agreement with the theoretical calculations (Fig.~3J) and shows that the optimal point for qubit operations is $n_g^\theta = 1/4$ where the Rabi frequency associated to both even and charge charge states coincide.

Having established the charge-dependent nature of the excited levels of the circuit and the optimal charge operation point of the qubit, we now turn to the population transfer between the protected ground states using the charge-insensitive $|\pi_{d\theta}^-\rangle$ ancillary state. First, to unambiguously demonstrate the existence of protected states, we again perform multi-tone spectroscopy between the lowest-lying states of the valleys ($|0_{s}\rangle$, $|\pi_{s}^+\rangle$, $|\pi_{s}^-\rangle$) and the ancillary level, which form a double $\Lambda$-configuration (Fig.~4A inset). By strongly driving the system near the $|\pi_{d\theta}^-\rangle\leftrightarrow|\pi_{s}^\pm\rangle$ transitions and probing $|0_{s}\rangle\leftrightarrow|\pi_{d\theta}^-\rangle$, we resolve two Autler-Townes doublets (Fig.~4A). These correspond to the dressed states associated with the fluxon transitions of the lowest-lying symmetric $|\pi_{s}^{+}\rangle$ and antisymmetric $|\pi_{s}^{-}\rangle$ states. This scheme enables us to map the excitations of the protected states without excessive drive amplitudes. At finite detunings from the ancillary level, we observe the signature of stimulated Raman transitions as a pair of lines with the slope of +1 when the frequency difference of the probe and coupler tones is on resonance with the transitions of $|0_{s}\rangle\leftrightarrow|\pi_{s}^{-}\rangle$ or $|0_{s}\rangle\leftrightarrow|\pi_{s}^{+}\rangle$. In the vicinity of $\Phi_\mathrm{ext}=0$ (Fig.~4B), the Raman transitions allow us to map out the hybridization gap formed between the lowest-lying states $|\pi_{s}^\pm\rangle$ of the $\theta = \pi$ valley . The spectroscopy data showcase a magnetic-flux sweet spot and a hybridization gap of $\Delta_H/2\pi\approx20$ MHz for the disjoint levels. This demonstrates that the $0 - \pi$ circuit harbors protected qubit states, which can easily be coupled to each other by an ancillary higher energy level.

We achieve coherent control of the qubit states using Raman gates via  $|\pi_{d\theta}^-\rangle$. To coherently transfer the population between the $|0_{s}\rangle$ and $|\pi_{s}^+\rangle$ ground states, we use two simultaneous Gaussian-shaped pulses with amplitudes $\Omega_\alpha$ and $\Omega_\beta$. The frequencies of the pulses are chosen to link the two protected ground states via the ancillary level, and have a detuning of $\Delta/2\pi$ from $|\pi_{d\theta}^-\rangle$ (Fig.~4C inset). In this Raman scheme, the two pulses and the truncated three-level system effectively exhibit two-level dynamics with only negligible occupation of the intermediate state\cite{steck,supplement}. In this way, we demonstrate Rabi oscillations between the protected ground states by first fixing the detuning of the pulses and independently varying the amplitudes of the two drives (Fig.~4C). In this protocol \cite{chimczak}, the largest population transfer can be realized when the two amplitudes are equal ($\Omega_\alpha=\Omega_\beta$). In Fig.~4D, we show coherent manipulation by keeping equal drive amplitudes ($\Omega_\alpha=\Omega_\beta=\Omega$) and varying the detuning $\Delta$ from the intermediate level, which results in oscillations in good agreement with an effective Rabi amplitude of $\Omega_R \propto \Omega^2/\Delta$ \cite{steck,supplement}. 

These time-domain measurements allow us to find the amplitudes for $\pi$ and $\pi$/2 pulses between the protected states to characterize the lifetime and coherence of the protected states. Energy relaxation measurements yield $T_1=1.56\pm0.1$ ms, which is an order of magnitude improvement over current state-of-the-art transmons \cite{dial} and comparable to the results reported on highly flux-sensitive heavy fluxonium \cite{earnest,lin}. Moreover, Ramsey interferometry yields $T_{2R}=8.5\pm0.6$ $\mu$s and Hahn echo measurement results in $T_{2E}=25.8\pm1.4$ $\mu$s at $\Phi_\mathrm{ext}=0$, which demonstrates first-order protection against flux noise and an order of magnitude improvement for the coherence times of qubits with disjoint suppport \cite{earnest,lin}. We anticipate that the coherence times and gate operations can be further improved with future designs by increasing the kinetic energy anisotropy (for instance by moving to a layered three-dimensional capacitor structure), reducing the susceptibility of the junctions to quasiparticle poisoning, and taking advantage of optimal control techniques \cite{abdelhafez}.  

Our work demonstrates the experimental realization of an intrinsically error-protected $0-\pi$ superconducting qubit, opening new avenues for robust encoding of quantum information in artificial atoms. The ability to engineer eigenstates of a qubit Hamiltonian with disjoint support yields prospects not only for the exploration of protected devices with superior coherence times, but also for simulation of solid state systems and the exploration of fundamental physical phenomena.

\begin{spacing}{0.5}
\bibliography{scibib}
\bibliographystyle{Science}
\end{spacing}

\section*{Acknowledgments}

We thank A. Vrajitoarea, P. Groszkowski, N. Earnest and A. Shearrow for helpful discussions. Work at Princeton, Northwestern and Chicago was supported by Army Research Office Grant No. W911NF-1910016. Devices were fabricated in the Princeton University Quantum
Device Nanofabrication Laboratory and in the Princeton Institute for the Science and Technology of Materials (PRISM) cleanroom. The authors acknowledge the use of Princeton's Imaging and Analysis Center, which is partially supported by the Princeton Center for Complex Materials, a National Science Foundation (NSF)-MRSEC program (DMR-1420541). This work was undertaken in part thanks to funding from NSERC and the Canada First Research Excellence Fund.

\section*{Supplementary materials}
Materials and Methods\\
Supplementary Text\\
Figs.~S1 to S7\\
Tables S1 to S4\\
References \cite{johansson,kohn}

\clearpage

\begin{figure}[htp]
    \centering
    \includegraphics[width=17cm]{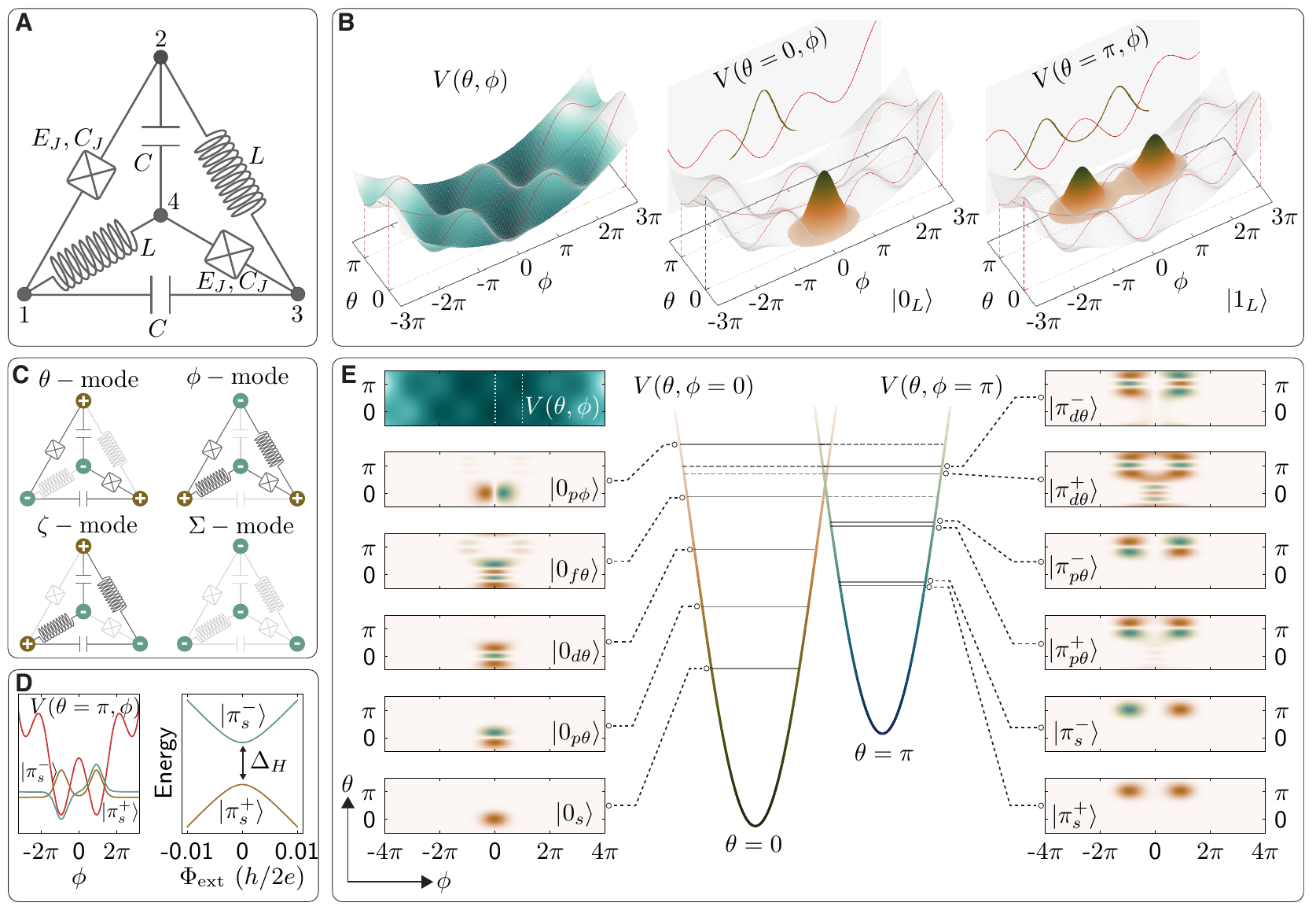}
\end{figure}
\begin{spacing}{1.0}
\noindent {\bf Fig.~1.} \textbf{The $\mathbf{0-\pi}$ superconducting qubit and its energy level structure.} (\textbf{A}) The circuit diagram of the $0-\pi$ qubit \cite{kitaev1,brooks}. The circuit has one closed loop with four nodes connected by a pair of Josephson junctions ($E_J, C_J$), large capacitors ($C$) and superinductors ($L$). (\textbf{B}) Left panel: the $V(\theta,\phi)$ double-well potential landscape of the circuit in the absence of magnetic fields. The ground state of the 0 valley is localized along $\theta=0$ (middle panel), while the lowest-lying state of the $\pi$ valley along $\theta=\pi$ (right panel). The line cuts along the two valleys in the $\phi$ direction show that the potential resembles a fluxonium potential. (\textbf{C}) The four modes of the $0-\pi$ circuit with colors of the nodes indicating the sign of normal-mode amplitudes. (\textbf{D}) Left panel: schematic of the symmetric and antisymmetric ground states of the $\pi$ valley. The hybridization of these states leads to a magnetic sweet spot (right panel). (\textbf{E}) The two-dimensional wavefunctions of the eigenstates, which are located mostly in the 0 (left) or in the $\pi$ (right) valleys. Middle panel: linecuts of the potential along $\phi=0$ and $\phi=\pi$ as indicated with white dotted lines on the image of the potential.     
\end{spacing}

\clearpage

\begin{figure}[htp]
    \centering
    \includegraphics[width=17cm]{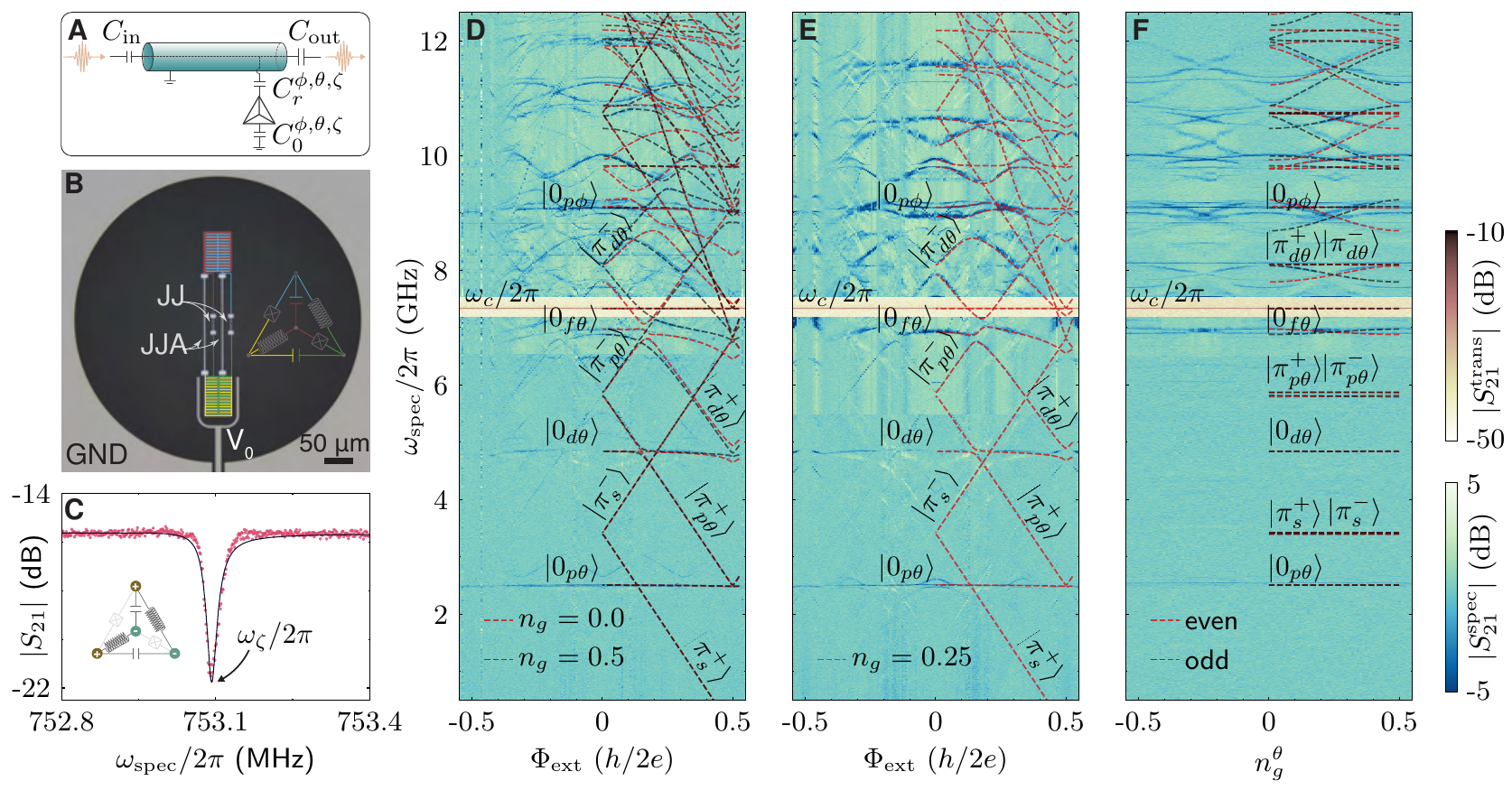}
\end{figure}
\begin{spacing}{1.0}
\noindent {\bf Fig.~2.} \textbf{Circuit QED with the $\mathbf{0-\pi}$ qubit.} (\textbf{A}) Schematic of the capacitive coupling scheme between the qubit and the transmission-line resonator. The capacitances between the four nodes of the circuit and the resonator determine the effective coupling capacitances for the modes. (\textbf{B}) False-color optical image of the ${0-\pi}$ device with colors referring to the four nodes of the circuit. GND: ground plane of the resonator; $\mathrm{V_0}$: centerpin of the resonator; JJ: Josephson junction; JJA: Josephson junction array. (\textbf{C}) The spectroscopic response of the harmonic $\zeta$ mode. Solid line shows the fit\cite{masluk} with quality factors of $Q_\mathrm{ext}^\zeta=41,600$ and $Q_\mathrm{int}^\zeta=42,500$. (\textbf{D} to \textbf{F}) Transmission and spectroscopy measurements (background subtracted) of the ${0-\pi}$ qubit as a function of external magnetic field (\textbf{D}) at $n_g^\theta=0.0/0.5$ and (\textbf{E}) at $n_g^\theta=0.25$, and as a function of offset-charge bias (\textbf{F}) at $\Phi_\mathrm{ext}=0$. The transmission measurements around 7.3 GHz (yellow-pink) show negligible dependence of the cavity resonance on external parameters. The spectroscopic data (green-blue) demonstrate the energy level structure of the ${0-\pi}$ qubit, which is in excellent agreement with a coupled resonator-qubit theoretical fit (dashed lines). The result of the fit is plotted over only the positive side of the data for clarity. The low-energy fluxon transitions are not visible in the spectroscopy data due to the small dipole elements.
\end{spacing}

\clearpage

\begin{figure}[htp]
    \centering
    \includegraphics[width=17cm]{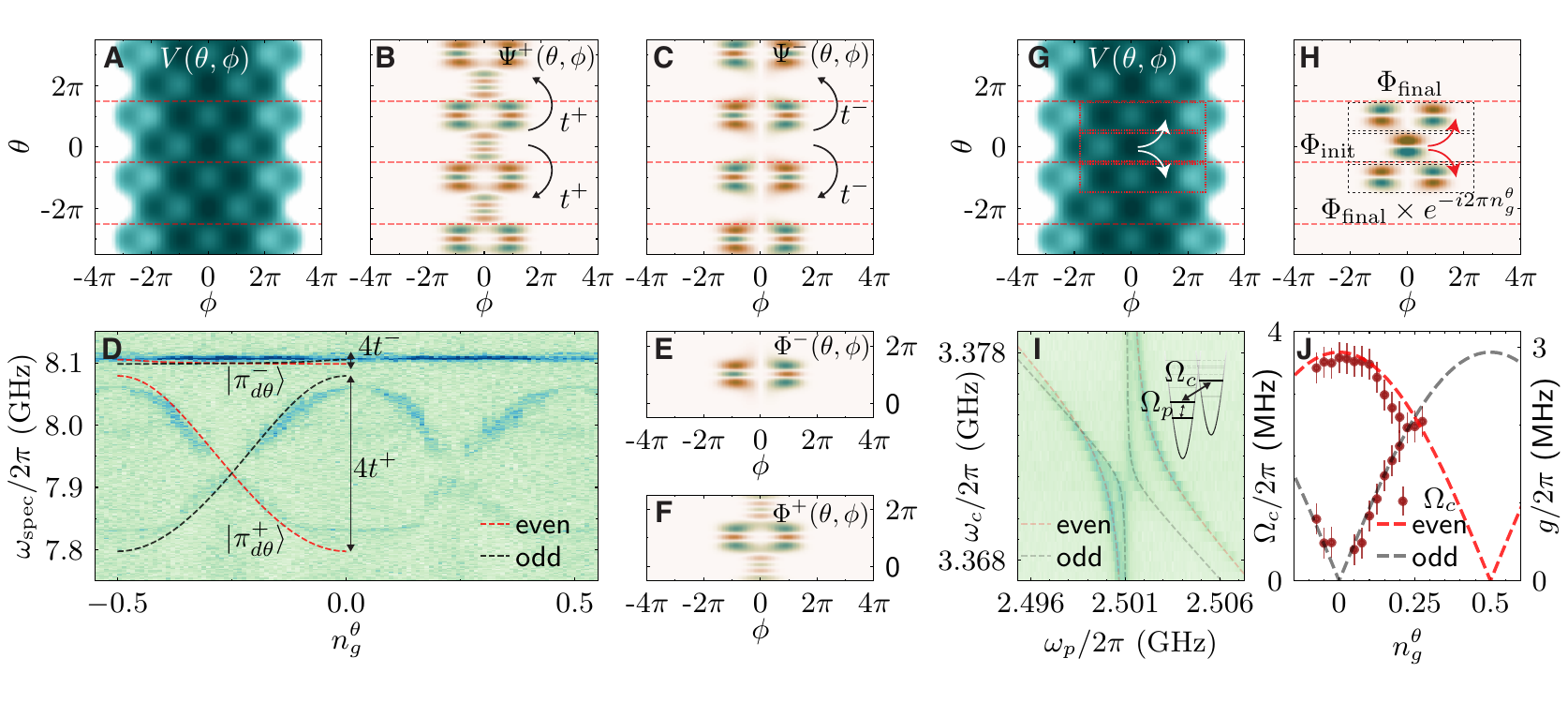}
\end{figure}
\begin{spacing}{1.0}
\noindent {\bf Fig.~3.} \textbf{Symmetry-induced offset-charge insensitivity and drive-assisted Aharonov-Casher effect.} (\textbf{A}) The extended potential of the ${0-\pi}$ circuit with red dashed lines indicating the border of the phase unit cells. (\textbf{B} and \textbf{C}) The eigenstates are Bloch-waves $\Psi(\theta,\phi)$, which are shown for the $|\pi_{d\theta}^\pm\rangle$ pair at $n_g^\theta=0.5$. (\textbf{E} and \textbf{F}) The antisymmetric Wannier wavefunction $\Phi^-(\theta,\phi)$ is significantly more localized than the symmetric state $\Phi^+(\theta,\phi)$, leading to a decreased hopping rate between adjacent unit cells: $t^-\ll t^+$. (\textbf{D}) Spectroscopic measurement of the charge dispersion of the $|\pi_{d\theta}^\pm\rangle$ pair (dashed lines show the same theoretical fit as in Fig.~2D to F). The antisymmetric state has suppressed charge sensitivity compared to the symmetric state in agreement with the different hopping rates. (\textbf{G}) Fluxon transition in the extended picture showing that a state located in the 0 valley can be excited to the valleys at $\pm\pi$. (\textbf{H}) Wannier function of the initial state $|0_{p\theta}\rangle$ located in the 0 valley and the final state $|\pi_{p\theta}^-\rangle$ located in the $\pi$ valley or in the $-\pi$ valley. The state in the $-\pi$ valley has a non-zero geometric phase due to the quasiperiodic boundary conditions \cite{supplement}. (\textbf{I}) Autler-Townes spectroscopy between $|0_{s}\rangle$, $|0_{p\theta}\rangle$ and $|\pi_{p\theta}^-\rangle$ at $n_g=0.1$ [dashed lines show the fit based on Rabi splitting of levels \cite{supplement}]. (\textbf{J}) The extracted Rabi splitting as a function of charge bias and the theoretically expected coupling rate $g/2\pi$ between the levels, which demonstrates the interference pattern with $|\cos{\pi n_g^\theta}|$ dependence. Error bars are estimates based on the linewidth of the transitions. 

\end{spacing}

\clearpage

\begin{figure}[htp]
    \centering
    \includegraphics[width=17cm]{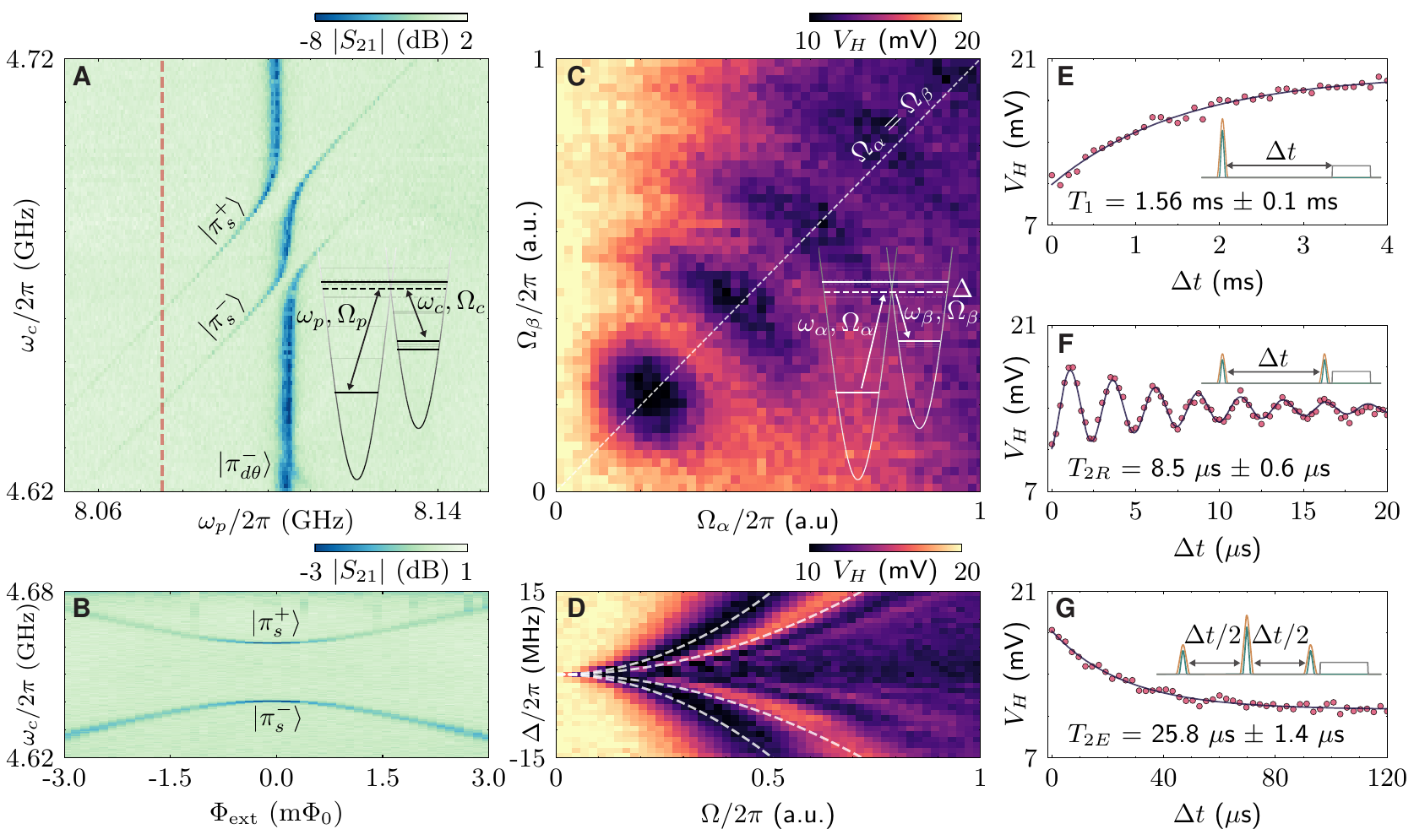}
\end{figure}
\begin{spacing}{1.0}
\noindent {\bf Fig.~4.} \textbf{Mapping and coherent control of protected quantum states.} (\textbf{A}) Autler-Townes spectroscopy (background subtracted) between the ground states of the $0-\pi$ qubit using the ancillary $|\pi_{d\theta}^-\rangle$ level. Inset shows the continuous drive scheme. (\textbf{B})  Raman spectroscopy at the detuning of $\Delta/2\pi=$ -30 MHz from the ancillary level [red dashed line in (A)] as a function of external magnetic field, which demonstrates a magnetic sweet spot for the disjoint ground states. (\textbf{C} and \textbf{D}) Coherent Rabi oscillations between the protected ground states $|0_s\rangle$ and $|\pi_s^{+}\rangle$ obtained by using two, overlapping Gaussian pulses with width of 4$\sigma$. (\textbf{C}) The measured homodyne voltage $V_H$ as a function of drive amplitudes at a fixed detuning ($\Delta/2\pi=$ -3 MHz, $\sigma=1$ $\mu$s), and (\textbf{D}) $V_H$ as a function of detuning at equal drive amplitudes ($\Omega_1=\Omega_2$, $\sigma=0.8$ $\mu$s). The maximum population transfer occurs when the two drive amplitudes are equal. The dashed lines in (D) show a fit according to the effective Rabi rate of the Raman pulses. (\textbf{E} to \textbf{G}) Relaxation, Ramsey and spin-echo measurements of the protected $|\pi_s^{+}\rangle$ state, with insets showing the pulse scheme ($\Delta/2\pi=$ -4 MHz  and $\sigma=$ 200 ns). All data were taken at $n_g^\theta=0.25$ charge bias point.

\end{spacing}
\end{document}


\begin{center}
\Large{Supplementary Materials for\\}
\vspace{5mm}
\large{Experimental realization of an intrinsically error-protected \\ superconducting qubit}
\\
\vspace{5mm}
\normalsize{Andr\'as Gyenis, Pranav S. Mundada, Agustin Di Paolo, Thomas M. Hazard,
\\
\normalsize Xinyuan You, David I. Schuster,
Jens Koch, Alexandre Blais, Andrew A. Houck$^{\ast}$}
\\
\vspace{5mm}
\small{$^{\ast}$Corresponding author. Email: aahouck@princeton.edu}
\end{center}

\section{Materials and Methods}
\subsection{Sample fabrication}

The device was fabricated on a 530 $\mu$m thick, polished c-plane sapphire substrate, on which 200 nm thick niobium was sputtered using an AJA superconducting deposition system. We used optical lithography to define the resonators and shunting capacitances. AZ1505 positive photoresist was spun on the chip, baked at 95$^{\circ}$C for 1 min and patterned using the 2 mm write-head of a Heidelberg DWL66+ tool. After developing the chip in AZ300MIF for 1 min and rinsed in running DI water for $\sim$1 min, the sample was dry-etched in PlasmaTherm APEX SLR using the mixture of CHF$_3$, O$_2$, SF$_6$, Ar gases (with 40:1:15:10 ratios). The photoresist was stripped by Microposit Remover 1165 and solvent-cleaned by toluene, acetone, methanol, isopropanol involving sonication and a nitrogen blow-dry. For electron-beam lithography, we span MMA/PMMA bilayer on the chip (baked for 2 + 30 min at 175$^{\circ}$C), evaporated 40 nm thick anticharging aluminum layer, and diced the sample into single chips. We exposed the Josephson junctions in a 125 keV Elionix e-beam system (at beam current of 1 nA and aperture of 60 $\mu$m). The anticharging layer was removed by soaking the chip in MF319 for 3 min and the e-beam resists were developed in the 1:3 mixture of methyl isobutyl ketone (MIBK) to isopropanol for 50 sec and pure isopropanol for 10 sec. The Josephson junctions were double-angle-evaporated in a Plassys e-beam-evaporator system with base pressure less than 10$^{-7}$ mbar. Before the evaporation, an \textit{in-situ} argon ion beam etch was used to clean the surface of the sample. We evaporated 20 nm + 50 nm thick Al layers at a rate of 0.4 nm/s and oxidized the first layer for 10 min at 200 mbar in a 15\% oxygen-in-argon environment to realize the tunnel junction. The Al layer was lift-off in PG Remover at $\sim$70$^{\circ}$C and cleaned with isopropanol. 

\begin{figure}[h!]
    \centering
    \includegraphics[width=17cm]{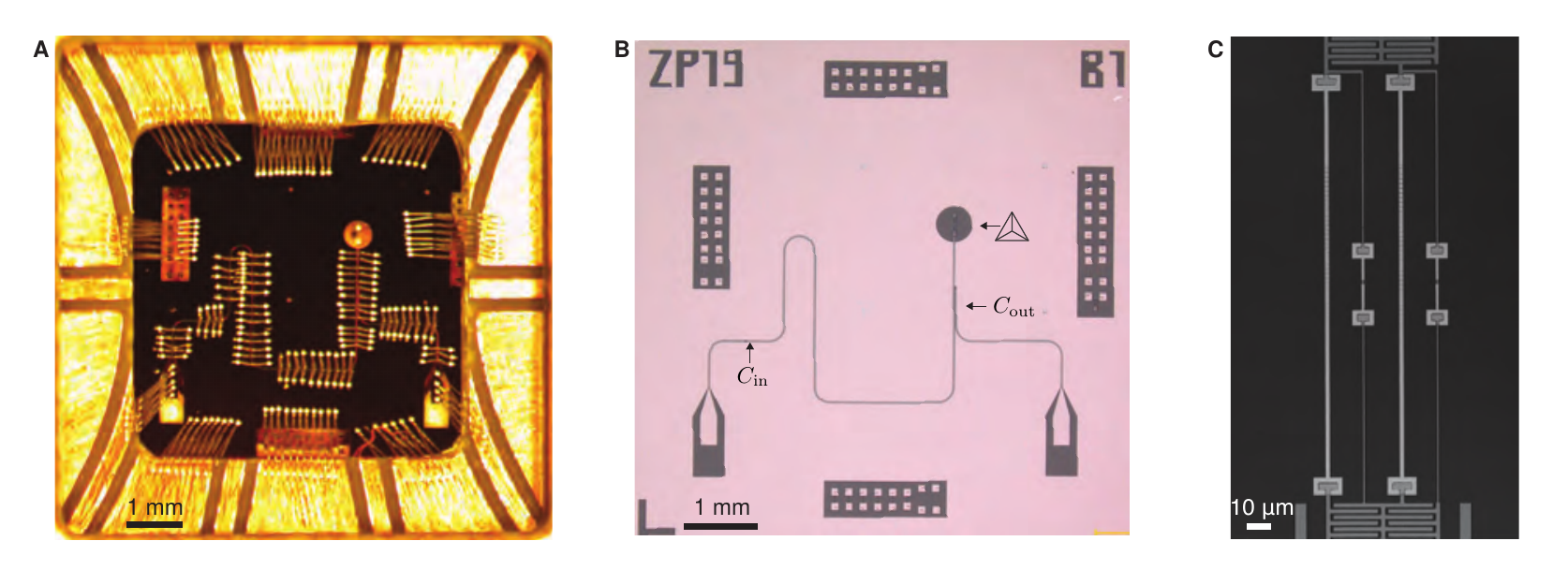}
    \caption{\label{fig:device_photo}(\textbf{A}) Optical image of a wire-bonded $0-\pi$ device mounted into the sample holder. (\textbf{B}) Image of the 7 mm x 7 mm chip showing the resonator with its coupling capacitors and the qubit. (\textbf{C}) Enlarged image of the middle region of the $0-\pi$ device, which displays the pairs of superinductors and Josephson junctions.}
\end{figure}

\begin{figure}[h!]
    \centering
    \includegraphics[width=17cm]{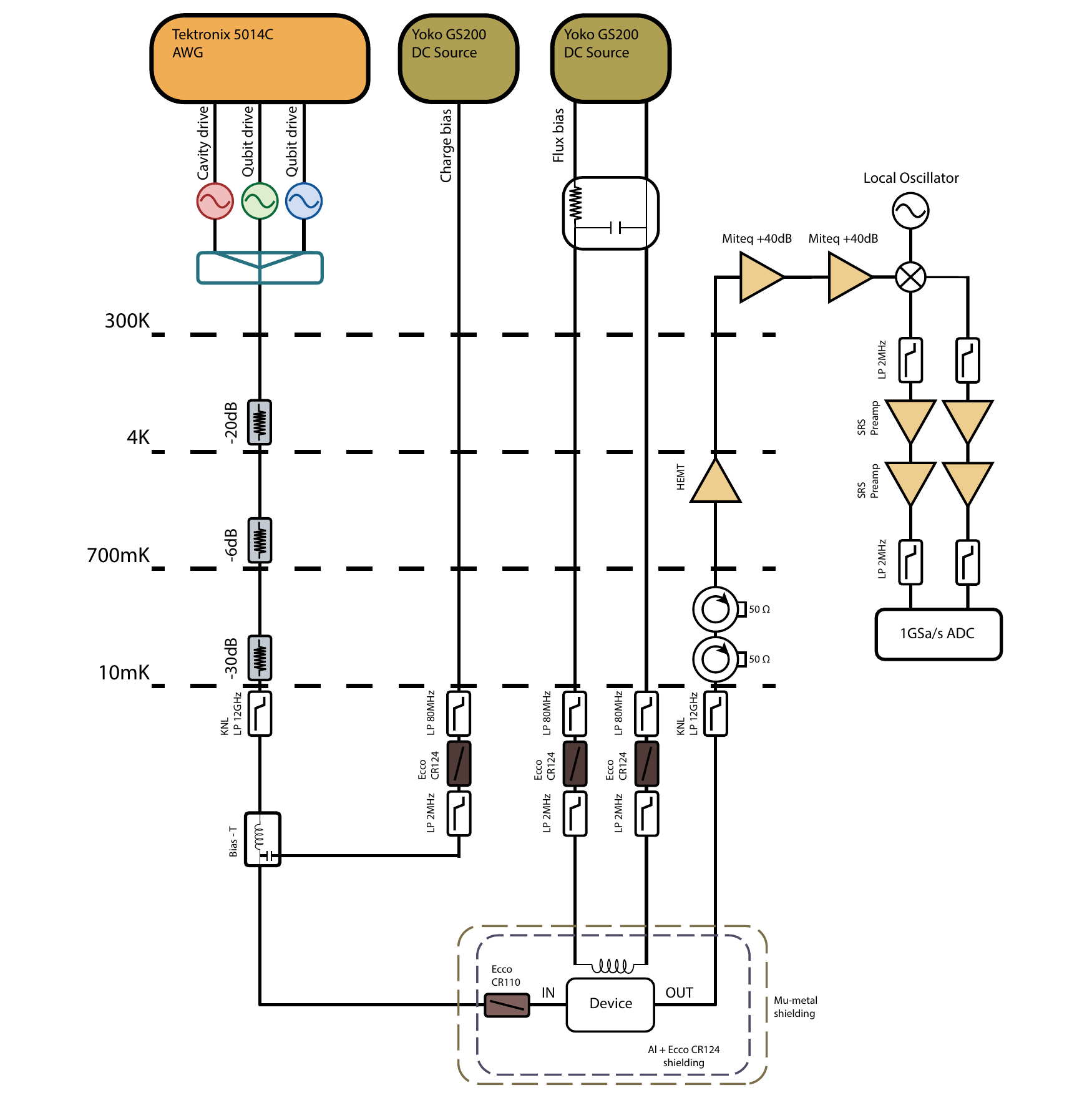}
    \caption{\label{fig:fridge_wiring} Wiring diagram of the cryogenic- and room-temperature measurement setup.}
\end{figure}

The device was placed in a copper PCB and wirebonded (\cref{fig:device_photo}). An off-chip copper coil was attached to the PCB.  The sample holder had an aluminum shield (covered with Eccosorb CR-124 and wrapped with thin Mylar layers) and an outer mu-metal shield. The sample holder was attached to the mixing chamber plate of a dilution refrigerator with base temperature of 10 mK (\cref{fig:fridge_wiring}).

\subsection{Finite-element simulation of the capacitances}

As mentioned in the main text, realizing the proper capacitance values in the $0-\pi$ circuit is a key requirement to achieve the protected regime. The large shunting capacitance in the circuit is denoted by $C$, while the cross capacitance between the nodes enclosing the superinductances (Josephson junctions) is $C_L^x$ ($C_J^x$). In our design, all four nodes are coupled to both the centerpin ($C_r^i$) and the ground plane ($C_0^i$) of the resonator (\cref{fig:capacitance_network}). We used ANSYS Maxwell electromagnetic field simulation software to determine the capacitance values in the circuit, which are summarized in \cref{tab:capacitance_parameters}. These parameters (with the assumptions of dielectric constant $\epsilon_r=10.7$ for sapphire, $C_J$ = 2 fF and $E_L = $ 0.38 GHz) results in energy scales of $E_C^\theta/h=$ 88 MHz, $E_C^\phi/h=$ 1.02 GHz and $\omega_\zeta/2\pi=$ 742 MHz, which are in excellent agreement with our experimental findings.

\begin{table}[h]
    \centering
    \begin{tabular}{ccccccccccc}
    $C$ & $C_L^x$ & $C_J^x$ & $C_r^1$ & $C_r^2$ & $C_r^3$ & $C_r^4$ & 
    $C_0^1$ & $C_0^2$ & $C_0^3$ & $C_0^4$\\\hline\hline
    100.5 & 0.7 & 1.0 & 9.1 & 0.3 & 3.8 & 0.3 & 8.2 & 7.9 & 6.2 & 11.6
    \end{tabular}
    \caption{\label{tab:capacitance_parameters} Finite-element simulation of the device capacitances.  All values are given in fF units.}
\end{table}

\clearpage

\clearpage

\subsection{Spectrum fit}
\label{s:Spectrum Fit}
\begin{figure}[h!]
    \centering
    \includegraphics[scale=1.5]{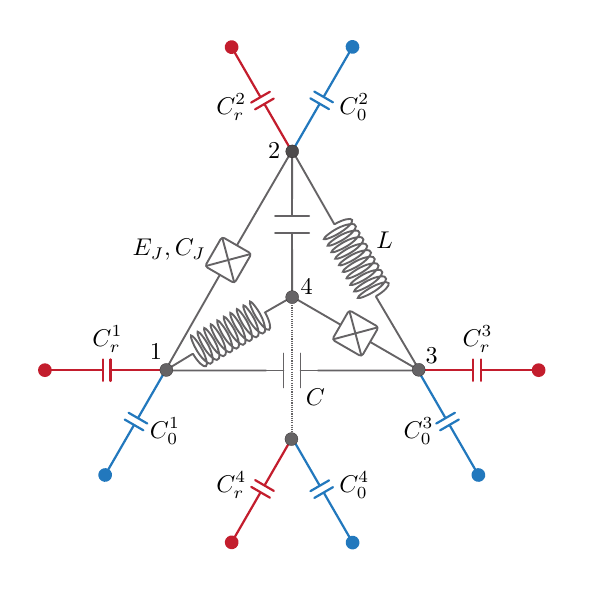}
    \caption{\label{fig:capacitance_network} Full capacitance network of the $0-\pi$ device. Red (blue) colors indicate coupling to the centerpin (ground plane) of the resonator.}
\end{figure}

Here we describe the multivariate fit to the experimental data based on a detailed theoretical model for the $0-\pi$ device. We consider the circuit scheme of \cref{fig:capacitance_network}, where we have introduced additional gate ($C_r^i$) and ground ($C_0^i$) capacitances for nodes $i\in[1,4]$. In the flux node basis $\{\Phi_i\}$, the circuit Lagrangian takes the form 
\begin{equation}
    \mathcal{L}_{\Phi} = \mathbf{\dot{\Phi}}^T\cdot\frac{\mathbf{C}_{\Phi}}{2}\cdot\mathbf{\dot{\Phi}} - \mathbf{\dot{\Phi}}^T\cdot\mathbf{C}_{r}\cdot\mathbf{V}_{\Phi} - U(\mathbf{\Phi},\Phi_{\mathrm{ext}}),
    \label{eq:circuit Lagrangian node basis}
\end{equation}
where $\mathbf{\Phi}=(\Phi_1,\dots,\Phi_4)^T$, $\mathbf{C}_{\Phi}$ is the capacitance matrix of the circuit (including gate and ground capacitances), $\mathbf{V}_{\Phi}= V_r (1,1,1,1)^T$ is a voltage-drive vector defined in terms of the resonator voltage, $V_r$, $\mathbf{C}_{r}=\mathrm{diag}(C_r^1,\dots,C_r^4)$ is the gate capacitance matrix, and $U(\mathbf{\Phi},\Phi_{\mathrm{ext}})$ is the potential energy corresponding to the Josephson junctions and inductances of the circuit. More precisely, the circuit capacitance matrix is given by
\begin{equation}
    \mathbf{C}_{\Phi}=
    \begin{pmatrix}
        C_1 & -C_J & -C & 0 \\
        -C_J & C_2 & 0 & -C \\
        -C & 0 & C_3 & -C_J \\
        0 & -C & -C_J & C_4 \\
    \end{pmatrix},
    \label{eq:full capacitance matrix}
\end{equation}
where $C_i = C_J + C + C_{r}^i + C_{0}^i$ for $i\in[1,4]$. We now move to the $0-\pi$ mode basis defined by $\mathbf{\Theta}=(\phi,\theta,\zeta,\Sigma)^T$, by the rotation $\mathbf{\Theta} = \mathbf{R}\cdot\mathbf{\Phi}$,  where 
\begin{equation}
    \mathbf{R}=\frac{1}{2}
    \begin{pmatrix}
        -1 & 1 & -1 & 1 \\
        -1 & 1 & 1 & -1 \\
        1 & 1 & -1 & -1 \\
        1 & 1 & 1 & 1\textbf{} \\
    \end{pmatrix}.
    \label{eq:coordinate transformation}
\end{equation}
Under such a transformation, \cref{eq:circuit Lagrangian node basis} becomes
\begin{equation}
    \mathcal{L}_{\Theta} = \mathbf{\dot{\Theta}}^T\cdot\frac{\mathbf{C}_{\Theta}}{2}\cdot\mathbf{\dot{\Theta}} - \mathbf{\dot{\Theta}}^T\cdot\mathbf{\tilde{C}}_{r}\cdot\mathbf{V}_{\Theta}- U(\mathbf{\Theta},\Phi_{\mathrm{ext}}),
    \label{eq:circuit Lagrangian mode basis}
\end{equation}
where $\mathbf{C}_{\Theta} = (\mathbf{R}^{-1})^T\cdot\mathbf{C}_{\Phi}\cdot\mathbf{R}^{-1}$ and $\mathbf{\tilde{C}}_{r} = (\mathbf{R}^{-1})^T\cdot\mathbf{{C}}_{r}\cdot\mathbf{R}^{-1}$ are the transformed capacitance matrices, and $\mathbf{V}_{\Theta} = \mathbf{R}\cdot\mathbf{V}_{\Phi}$ is the voltage-drive vector expressed in the $0-\pi$ mode basis. By performing a Legendre transformation, we arrive at the circuit Hamiltonian
\begin{equation}
    H = (\mathbf{q}_{\Theta} + \mathbf{\tilde{C}}_{r}\cdot\mathbf{V}_{\Theta})^T\cdot\frac{\mathbf{C}_{\Theta}^{-1}}{2}\cdot(\mathbf{q}_{\Theta} + \mathbf{\tilde{C}}_{r}\cdot\mathbf{V}_{\Theta}) + U(\mathbf{\Theta},\Phi_{\mathrm{ext}}),
    \label{eq:circuit Hamiltonian mode basis}
\end{equation}
where $\mathbf{q}_{\Theta} = {\partial{\mathcal{L}_{\Theta}}}/{{\partial \mathbf{\dot{\Theta}}}}$ is the conjugate charge vector operator. Note that \cref{eq:circuit Hamiltonian mode basis} can be split as
\begin{equation}
    H = H_{0-\pi} + H_{\mathrm{drive}},
    \label{eq:circuit Hamiltonian mode basis v2}
\end{equation}
where 
\begin{equation}
    H_{0-\pi}= \mathbf{q}_{\Theta}^T\cdot\frac{\mathbf{C}_{\Theta}^{-1}}{2}\cdot\mathbf{q}_{\Theta} + U(\mathbf{\Theta},\Phi_{\mathrm{ext}}),
    \label{eq:circuit Hamiltonian mode basis ZP only}
\end{equation}
is the undriven $0-\pi$ qubit Hamiltonian and 
\begin{equation}
    H_{\mathrm{drive}} = \mathbf{q}_{\Theta}^T\cdot(\mathbf{C}_{\Theta}^{-1}\cdot\mathbf{\tilde{C}}_{r})\cdot\mathbf{V}_{\Theta},
    \label{eq:circuit Hamiltonian mode basis drive only}
\end{equation}
is the drive term. 

While all circuit details are taken into account in \cref{eq:circuit Hamiltonian mode basis v2}, the spectrum fit that is presented in the main text aims to provide the simplest possible accurate description of the device Hamiltonian. Thus, in order to simplify our treatment, we implement a few approximations. In particular, we omit any coupling to the $\zeta$ and $\Sigma$ modes, neglecting a potential capacitive interaction between these and the qubit modes and reducing the qubit Hamiltonian to 
\begin{equation}
    H_{0-\pi} \simeq 4E_{C}^{\phi}n_{\phi}^2 + 4E_{C}^{\theta}(n_{\theta}-n_g)^2 + \hbar g_{\phi\theta} n_{\phi}n_{\theta} + U(\mathbf{\Theta},\Phi_{\mathrm{ext}}).
    \label{eq:circuit Hamiltonian mode basis ZP only reduced}
\end{equation}
Here, $E_{C}^{\phi}=e^2/2C_{\phi}$ and $E_{C}^{\theta}=e^2/2C_{\theta}$ are the charging energies of the $\phi$ and $\theta$ modes and $\hbar g_{\phi\theta}$ is the strength of a capacitive interaction between these modes due to the asymmetry of the circuit capacitance matrix. Accordingly, we also approximate \cref{eq:circuit Hamiltonian mode basis drive only} by
\begin{equation}
    H_{\mathrm{drive}} \simeq (\beta_{\phi}n_{\phi} + \beta_{\theta}n_{\theta})\times2eV_{r},
    \label{eq:circuit Hamiltonian mode basis drive only approx}
\end{equation}
where $\beta_{\phi}$ and $\beta_{\theta}$ are capacitive coupling ratios for the $\phi$ and $\theta$ modes. We moreover set $g_{\phi\theta}\to 0$ in \cref{eq:circuit Hamiltonian mode basis drive only approx}, eliminating one fit parameter. We observed, however, that deviations from $g_{\phi\theta}\simeq 0$ within bounds given by finite-element estimations of the coupling capacitance do not significantly modify the quality of the fit. 

\begin{figure}[h!]
    \centering
    \includegraphics[width=17cm]{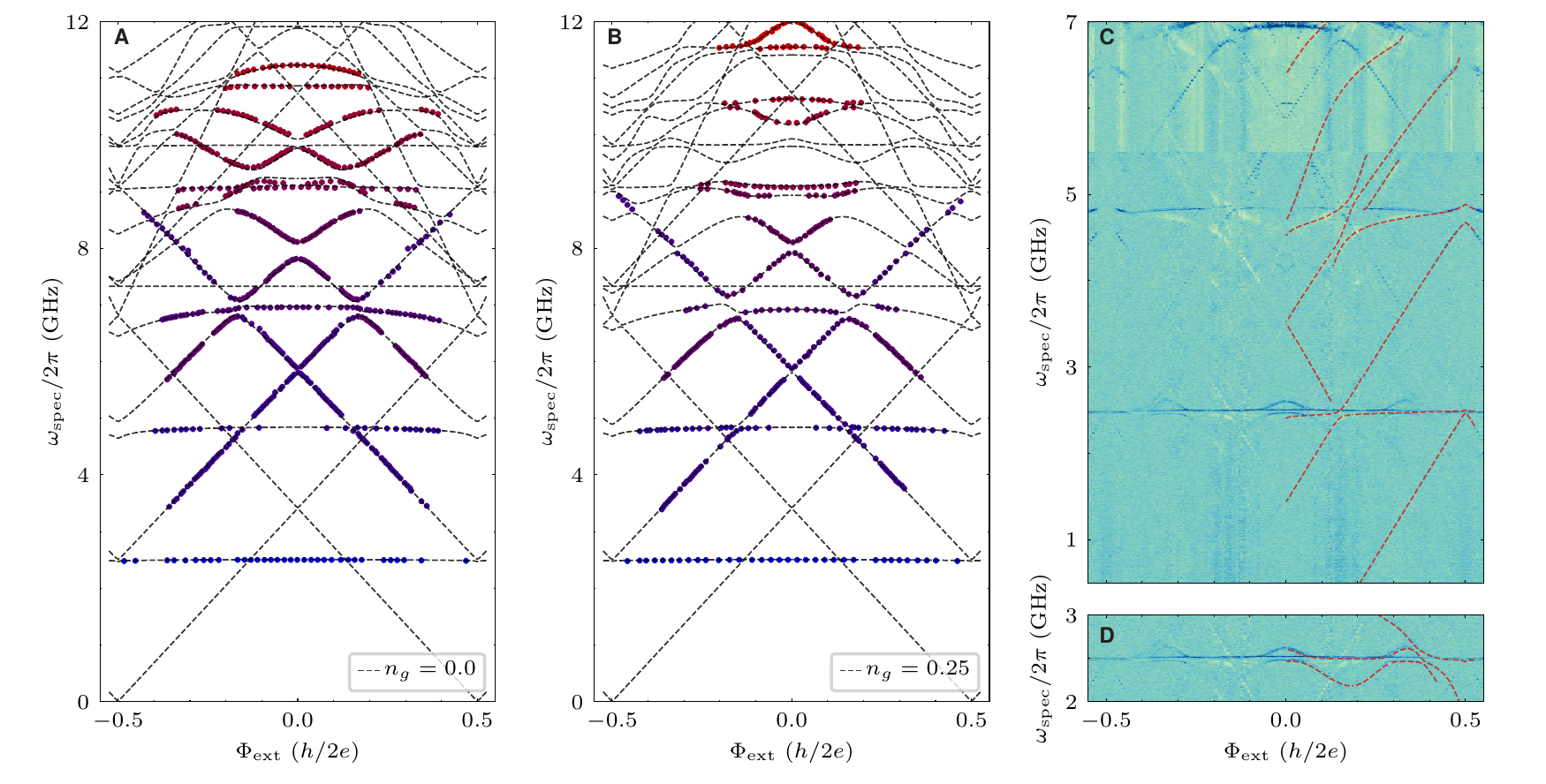}
    \caption{\label{fig:spectrum_details} Spectrum fit of the $0-\pi$ device. (\textbf{A}) and (\textbf{B}) correspond to the cases of $n_g=0$ and $n_g=0.25$, respectively. The experimental data used for the fit are displayed by colored circles while the theory results are given by the black dashed lines. Note that some of the qubit transitions are invisible in the experiment due to exponentially small matrix elements or vanishing dispersive shifts. The transition around $7.35\,\mathrm{GHz}$ corresponds to the readout resonator. (\textbf{C}) Overlay of the transitions predicted by our model on the experimental data by assuming thermal population of the lowest two levels in the $\pi$ valley, which explains the origin of the \textit{bright} transitions in the spectrum ($n_g=0.25$). (\textbf{D}) A set of cavity-assited sideband transtions that are also captured by the theory model ($n_g=0.25$).}
\end{figure}

For the multivariate fit, we treat all energy and coupling variables as fit parameters, including $E_{C_{\phi}}$, $E_{C_{\theta}}$, $\beta_{\phi}$, $\beta_{\theta}$ and those in the potential energy
\begin{equation}
    U(\mathbf{\Theta},\Phi_{\mathrm{ext}}) = -2E_J\cos\theta\cos(\phi-{\pi\Phi_{\mathrm{ext}}}/{\Phi_0}) + E_L\phi^2 + E_J dE_J \sin\theta\sin(\phi-{\pi\Phi_{\mathrm{ext}}}/{\Phi_0}),
    \label{eq:potential energy}
\end{equation}
defined in terms of the junction energy $E_J$, the superinductance energy $E_L$ and the relative junction-energy asymmetry $dE_J$. The fit also incorporates the resonator mode with nominal impedance $Z_r=50\,\Omega$ and frequency $f_r\simeq 7.35\,\mathrm{GHz}$ parameters, for which the voltage operator reads
\begin{equation}
    V_r= V_{\mathrm{rms}} (a + a^{\dagger}),
    \label{eq:Vrms}
\end{equation}
where $V_{\mathrm{rms}} = \sqrt{2hf_r^2Z_r}$ for a $\lambda/2$ resonator, and $a$ and $a^{\dagger}$ are the respective harmonic-oscillator ladder operators. The fit takes into account two sets of data corresponding to a sweep of the magnetic flux for the offset charges $n_g=0.0$ and $n_g=0.25$. A single error metric measures the distance between the result of the exact diagonalization of the qubit-resonator Hamiltonian and both data sets. The result of the fit is shown in \cref{fig:spectrum_details}A and B (in addition to the figures in the main manuscript)
and the fit parameters are provided in \cref{tab:fit_parameters}. These parameters are in excellent agreement with those expected from a finite-element simulation of the device.
\begin{table}[h]
    \centering
    \begin{tabular}{ccccccc}
    $E_{C}^{\phi}/h$ & $E_{C}^{\theta}/h$ & $E_J/h$ & $E_L/h$ & $dE_J$ & $\beta_{\phi}$ & $\beta_{\theta}$ \\\hline\hline
    1.142 & 0.092 & 6.013 & 0.377 & 0.1 & $0.27$ & $6.6\times 10^{-3}$
    \end{tabular}
    \caption{\label{tab:fit_parameters} Result of the multivariate fit to the experimental data. All energy parameters are given in GHz units.}
\end{table}

We find an excellent agreement between the theoretical model and the experimental data, both for $n_g=0$ and $n_g=0.25$. As Fig.~2 of the main text shows, the obtained parameters also describe the transitions at $n_g=0.5$, and generally, the entire charge dependence of the levels. In  \cref{fig:spectrum_details}C, we also show that additional features in the spectroscopy data can be explained by transitions between the thermally occupied fluxon states to higher levels.  Furthermore, the theoretical model not only captures accurately the qubit transitions, but also the cavity-assited sideband transitions. Since the latter transitions were not originally taken into account for the fit, this fact provides further confirmation of the validity of the theoretical model. 

\section{Supplementary text}

\subsection{Tight-binding approximation}

In the main text, we introduced a tight-binding model to explain the charge dependence of the fluxon transitions and dipole matrix elements. Here, we provide additional information regarding this model. We present a bottom-up approach considering first the case of a charge-sensitive transmon, after which we focus on the case of the $0-\pi$ qubit.

To make the connection between a charge-sensitive qubit and a periodic lattice, we briefly review the definitions of Bloch states and Wannier states in solid-state physics. The single-particle Hamiltonian describing electrons moving in a one-dimensional crystal with a periodic potential is
\begin{equation}\label{crystal_Hamiltonian}
    H_\mathrm{crystal}= -\frac{\hbar^2}{2m}\partial_x^2 + V(x),
\end{equation}
where $m$ is the mass of the electron and $\hbar$ is the reduced Planck constant. The eigenstates belonging to a single band are quasi-periodic Bloch states $\Psi_k(x) = e^{ikx} u_{k}(x)$, where $u_{k}(x)$ is a lattice periodic function and $k$ is the crystal-momentum. 

We define the Wannier function corresponding to a molecular orbital at a lattice site $x_0$ by 
\begin{equation}
    \Phi(x-x_0) = \frac{1}{\sqrt{N}}\sum_{k}e^{-ikx_0}\Psi_{k}(x), 
\end{equation}
where $N$ is the number of sites in the lattice. An advantage of using Wannier functions is that they provide a natural choice for localized, orthonormal atomic states. The Wannier functions are non-unique due to the unconstrained phase degree of freedom of the Bloch electrons. However, there exists only one maximally localized Wannier wavefunction that is real, exponentially localized and symmetric or antisymmetric (\textit{48}).

We can express the Bloch states as a function of the Wannier states corresponding to different lattice sites by an inverse Fourier transform
\begin{equation}
  \Psi_k(x) = \frac{1}{\sqrt{N}}\sum_{x_0} e^{ikx_0}\Phi(x-x_0).  
\end{equation}

The energy of the eigenstates as a function of momentum can be approximated by the well-known tight-binding dispersion relation of $\epsilon_k = \epsilon_0 + 2t\cos{ka}$, where the energy scales are related to the localized Wannier functions as $\epsilon_0 = \int{dx}\Phi^*(x)H\Phi(x)$ and $t = \int{dx}\Phi^*(x-a)H\Phi(x)$.

\paragraph{Charge-sensitive transmon}

We now consider the case of the offset-charge sensitive transmon qubit
\begin{equation}
    H= 4E_C(i\partial_\theta-n_g)^2-E_J\cos\theta,
\end{equation}
where $E_C$ is the charging energy and $E_J$ is the Josephson energy. The $p$th eigenstate of this Hamiltonian for a given $n_g$ obeys 
\begin{equation}
Hu_{n_g}^p(\theta)=\epsilon_{n_g}^p u_{n_g}^p(\theta),
\end{equation}
where $u_{n_g}^p(\theta)$ is a $2\pi$-periodic function of the superconducting phase across the Josephson junction, i.e.,
\begin{equation}
    u^p_{n_g}(\theta) = u^p_{n_g}(\theta + 2\pi).
\end{equation}
Next, we perform a gauge transformation defined by the unitary  $U=e^{in_g\theta}$, in order to eliminate the offset-charge dependence of the transmon Hamiltonian which becomes
\begin{equation}\label{transmon_Hamiltonian_periodic}
    \bar{H}= -4E_C\partial_\theta^2-E_J\cos\theta,
\end{equation}
and thus
\begin{equation}\label{transmon_wavefunction_quasiperiodic}
    \Psi^p_{n_g}(\theta) = e^{in_g\theta} u^p_{n_g}(\theta).
\end{equation}
We note that, in this gauge, the transmon Hamiltonian [\cref{transmon_Hamiltonian_periodic}] is identical to the Hamiltonian of a one-dimensional crystal [\cref{crystal_Hamiltonian}]. Therefore, the eigenstates of \cref{transmon_Hamiltonian_periodic} are also quasi-periodic Bloch waves.

In analogy to the solid state case, we introduce the Wannier functions for the transmon qubit by Fourier transforming the Bloch states in $n_g$
\begin{equation}
    \Phi_p(\theta-2\pi l) = \frac{1}{\sqrt{N}}\sum_{n_g}e^{-i2\pi l\cdot n_g}\Psi^p_{n_g}(\theta), 
\end{equation}
where $l$ is an integer that corresponds to the number of the unit cell where the Wannier function is localized. The inverse Fourier transform thus reads
\begin{equation}
    \Psi_{n_g}^p(\theta) = \frac{1}{\sqrt{N}}\sum_{\theta_0}e^{i2\pi l\cdot n_g}\Phi_p(\theta-2\pi l). 
\end{equation}

The $n_g$ dispersion of the low-lying transmon energy levels is determined by the hopping matrix element: $\epsilon_{n_g}^p = \epsilon_0^p + 2t^p\cos{2\pi n_g}$, where $\epsilon_0^p = \int{d\theta}\Phi^{p*}(\theta)H\Phi^p(\theta)$ and $t^p = \int{d\theta}\Phi^{p*}(\theta-2\pi)H\Phi^p(\theta)$.

Next, we consider the charge matrix element $n_{pq} = \langle u^p | i\partial_\theta | u^q \rangle$ in the tight-binding approximation. Since the states are assumed to be localized, we only consider the
contribution of two states which are in the same well or nearest neighbours, thus
\begin{align}
    \langle u^p | i\partial_\theta | u^q \rangle =\, & i\int d\theta \Phi_p^\ast(\theta) \partial_\theta \big(\Phi_q(\theta) + \Phi_q(\theta+2\pi)e^{-i 2\pi n_\text{g}} + \Phi_q(\theta-2\pi)e^{i 2\pi n_\text{g}}) \big)\notag \\
    & + n_\text{g}\int d\theta \Phi_p^\ast(\theta)\big(\Phi_q(\theta) + \Phi_q(\theta+2\pi)e^{-i 2\pi n_\text{g}} + \Phi_q(\theta-2\pi)e^{i 2\pi n_\text{g}} \big),
\end{align}
and for further reference, we define the following variables
\begin{gather*}
    \eta_0^\text{C} = \int d\theta \Phi_p^\ast(\theta) \Phi_q(\theta), \qquad \eta_0^\text{L} = \int d\theta \Phi_p^\ast(\theta) \Phi_q(\theta+2\pi), \\ \eta_0^\text{R} = \int d\theta \Phi_p^\ast(\theta) \Phi_q(\theta-2\pi), \qquad
    \eta_1^\text{C} = i\int d\theta \Phi_p^\ast(\theta)\partial_\theta \Phi_q(\theta), \\
    \eta_1^\text{L} = i\int d\theta \Phi_p^\ast(\theta)\partial_\theta \Phi_q(\theta+2\pi), \qquad \eta_1^\text{R} = i\int d\theta \Phi_p^\ast(\theta)\partial_\theta \Phi_q(\theta-2\pi).
\end{gather*}

The matrix element can be written as 
\begin{equation}
    \langle u^p | i\partial_\theta | u^q \rangle =\, (\eta_1^\text{C} +\eta_1^\text{L}e^{-i 2\pi n_\text{g}} + \eta_1^\text{R}e^{i 2\pi n_\text{g}} ) + n_\text{g}(\eta_0^\text{C} +\eta_0^\text{L}e^{-i 2\pi n_\text{g}} + \eta_0^\text{R}e^{i 2\pi n_\text{g}}).
\end{equation}
We now consider the case where $\Phi_p$ and $\Phi_q$ have the same parity. It follows then, that $\eta_0^\text{C}=\eta_1^\text{C}=0$, $\eta_1^\text{L}=-\eta_1^\text{R}$, and $\eta_0^\text{L}=\eta_0^\text{R}$. The matrix element thus simplifies to 
\begin{equation}
    |\langle u^p | i\partial_\theta | u^q \rangle| = |-2 i \eta_1^\text{L}\sin{2\pi n_\text{g}} + 2n_\text{g}\eta_0^\text{L}\cos{2\pi n_\text{g}}|.
\end{equation}
Since the states are all localized in the corresponding wells, the part that contributes to the integral is the tail of the wavefunction. Assuming the tail is of Gaussian type $\exp(-\theta^2)$, we have $|\eta_1^\text{L}|\gg|\eta_0^\text{L}|$, and the matrix element is further simplified 
\begin{equation}
    |\langle u^p | i\partial_\theta | u^q \rangle| = |2\eta_1^\text{L}\sin{2\pi n_\text{g}}|.
\end{equation}

The case in which $\Phi_p$ and $\Phi_q$ have opposite parities leads to $\eta_0^\text{C}=0$, $\eta_1^\text{L}=\eta_1^\text{R}$, and $\eta_0^\text{L}=-\eta_0^\text{R}$. The matrix element is then
\begin{equation}
    |\langle u^p | i\partial_\theta | u^q \rangle| = |\eta_1^\text{C} + 2\eta_1^\text{L}\cos{2\pi n_\text{g}} - 2 i n_\text{g}\eta_0^\text{L}\sin{2\pi n_\text{g}}|.
\end{equation}
Taking into account $|\eta_1^\text{C}|\gg|\eta_1^\text{L}|\gg|\eta_0^\text{L}|$, we have 
\begin{equation}
    |\langle u^p | i\partial_\theta | u^q \rangle| = |\eta_1^\text{C} + 2\eta_1^\text{L}\cos{2\pi n_\text{g}}|.
\end{equation}

\paragraph{Charge-sensitivity in the $0-\pi$ qubit}

The charge sensitivity in the $0-\pi$ qubit enters through the $\theta$ mode as
\begin{align}
    H_{0-\pi}=4E_C^\theta(n_\theta - n_g^\theta)^2 +4E_C^\phi n_\phi^2 + V(\theta,\phi).
\end{align}
The $u_p(\theta,\phi)$ eigenstates of the $0-\pi$ Hamiltonian are $2\pi$-periodic in $\theta$ as $u_p(\theta,\phi)=u_p(\theta+2\pi,\phi)$. We define the Bloch states again as $\Psi^p_{n_g}(\theta,\phi) = e^{in_g\theta} u_p(\theta,\phi)$ and the Wannier states as $\Phi_p(\theta-2\pi l,\phi) = \frac{1}{\sqrt{N}}\sum_{n_g}e^{-i2\pi l\cdot n_g}\Psi^p_{n_g}(\theta,\phi)$, where $l$ is an integer.

The low-energy levels are either localized in the 0 or in the $\pi$-valley, which we here explicitly note by a superscript: $u_p^0(\theta,\phi)$ and $u_p^\pi(\theta,\phi)$. We also distinguish between Wannier wavefunctions localized in the two different valleys by introducing 
\begin{align}
    \Phi_p^0(\theta,\phi)= \Phi_p(\theta,\phi), \\
    \Phi_p^\pi(\theta,\phi)= \Phi_p(\theta+\pi,\phi),
\end{align}
which are centered around the center of the 0 and $\pi$ valley, respectively.

Next we consider the charge matrix elements $\langle u_p^0 | i\partial_\theta | u_q^\pi \rangle$ and $\langle u_p^0 | i\partial_\phi | u_q^\pi \rangle$. Since the states are assumed to be localized either in the $\theta=0$ or $\theta=\pi$ well, we only consider the contribution of two states being the nearest neighbours, i.e. for $l$ component of $u_p^0$, we only keep $l$, $l+1$ components of $u_q^\pi$. The matrix elements are then
\begin{align}
    \langle u_p^0 | i\partial_\theta | u_q^\pi \rangle =\, & i\int d\theta d\phi\, \Phi_p^0(\theta,\phi) \partial_\theta \big(\Phi_q^\pi(\theta-\pi,\phi) + \Phi_q^\pi(\theta+\pi,\phi)e^{-i 2\pi n_\text{g}}  \big) \notag \\ 
    & + n_\text{g}\int d\theta d\phi \, \Phi_p^0(\theta,\phi)\big(\Phi_q^\pi(\theta-\pi,\phi) + \Phi_q^\pi(\theta+\pi,\phi)e^{-i 2\pi n_\text{g}}  \big), \\
    \langle u_p^0 | i\partial_\phi | u_q^\pi \rangle =\, & i\int d\theta d\phi\, \Phi_p^0(\theta,\phi) \partial_\phi \big(\Phi_q^\pi(\theta-\pi,\phi) + \Phi_q^\pi(\theta+\pi,\phi)e^{-i 2\pi n_\text{g}}  \big).
\end{align}
For simplicity, we define the following variables
\begin{align*}
    \eta_0^\text{L} &= \int d\theta d\phi \Phi_p^0(\theta,\phi) \Phi_q^\pi(\theta+\pi,\phi), \qquad \eta_0^\text{R} = \int d\theta d\phi  \Phi_p^0(\theta,\phi) \Phi_q^\pi(\theta-\pi,\phi),\\
    \eta_1^\text{L} &= i\int d\theta d\phi \Phi_p^0(\theta,\phi)\partial_\theta \Phi_q^\pi(\theta+\pi,\phi), \qquad \eta_1^\text{R} = i\int d\theta d\phi \Phi_p^0(\theta,\phi)\partial_\theta \Phi_q^\pi(\theta-\pi,\phi), \\
    \eta^\text{L} &= i\int d\theta d\phi \Phi_p^0(\theta,\phi)\partial_\phi \Phi_q^\pi(\theta+\pi,\phi), \qquad \eta^\text{R} = i\int d\theta d\phi \Phi_p^0(\theta,\phi)\partial_\phi \Phi_q^\pi(\theta-\pi,\phi). 
\end{align*}
The matrix element thus can be written as 
\begin{align}
    \langle u_p^0 | i\partial_\theta | u_q^\pi \rangle &=\, (\eta_1^\text{L}e^{-i 2\pi n_\text{g}} +\eta_1^\text{R}) + n_\text{g}(\eta_0^\text{L}e^{-i 2\pi n_\text{g}} +\eta_0^\text{R}), \\
    \langle u_p^0 | i\partial_\phi | u_q^\pi \rangle &=\, \eta^\text{L}e^{-i 2\pi n_\text{g}}  +\eta^\text{R} .
\end{align}

The dipole matrix elements can be further simplified depending on the parities along the $\theta$ and $\phi$ directions, which eventually leads to the charge dependence of the fluxon dipole matrix elements summarized in \cref{tab:fluxon_transitions}. The plasmon transition in the $0-\pi$ qubit is similar to the case of the transmon qubit, and the result is summarized in the \cref{tab:plasmon_transitions}.

\begin{table}[h]
\centering
\begin{tabular}{ |c|c|c|c|c| } 
 \hline
  & \thead{$\Pi_i^\theta = \Pi_j^\theta$, \\ $\Pi_i^\phi = \Pi_j^\phi $ } & \thead{$\Pi_i^\theta =- \Pi_j^\theta$, \\ $\Pi_i^\phi = \Pi_j^\phi $} & \thead{$\Pi_i^\theta = \Pi_j^\theta$,\\ $\Pi_i^\phi =- \Pi_j^\phi $} & \thead{$\Pi_i^\theta = -\Pi_j^\theta$, \\ $\Pi_i^\phi = -\Pi_j^\phi $}  \\ 
  \hline
  $|\langle i |i\partial_\theta | j \rangle|$ & $|\sin{\pi n_\text{g}}|$ & $|\cos{\pi n_\text{g}}|$ & 0 & 0   \\ 
  \hline
  $|\langle i |i\partial_\phi | j \rangle|$ & 0 & 0 & $|\cos{\pi n_\text{g}}|$ & $|\sin{\pi n_\text{g}}|$ \\ 
 \hline
\end{tabular}
\caption{\label{tab:fluxon_transitions}Matrix elements for fluxon transition.}
\label{table2}
\end{table}

\begin{table}[h]
\centering
\begin{tabular}{ |c|c|c|c|c| } 
 \hline
  & \thead{$\Pi_i^\theta = \Pi_j^\theta$, \\ $\Pi_i^\phi = \Pi_j^\phi $}  & \thead{$\Pi_i^\theta =- \Pi_j^\theta$, \\ $\Pi_i^\phi = \Pi_j^\phi $} & \thead{$\Pi_i^\theta = \Pi_j^\theta$, \\ $\Pi_i^\phi =- \Pi_j^\phi $} & \thead{$\Pi_i^\theta = -\Pi_j^\theta$, \\ $\Pi_i^\phi = -\Pi_j^\phi $}  \\ 
  \hline
  $|\langle i |i\partial_\theta | j \rangle|$ & $|\sin{2\pi n_\text{g}}|$ & $|1+\epsilon\cos{2\pi n_\text{g}}|$ & 0 & 0   \\ 
  \hline
  $|\langle i |i\partial_\phi | j \rangle|$ & 0 & 0 & $|1+\epsilon\cos{2\pi n_\text{g}}|$ & $|\sin{2\pi n_\text{g}}|$ \\ 
 \hline
\end{tabular}
\caption{\label{tab:plasmon_transitions}Matrix elements for plasmon transition.}
\label{table1}
\end{table}

\subsection{Population transfer in the two-tone Raman pulse scheme}
\begin{figure}[h!]
    \centering
    \includegraphics[width=17cm]{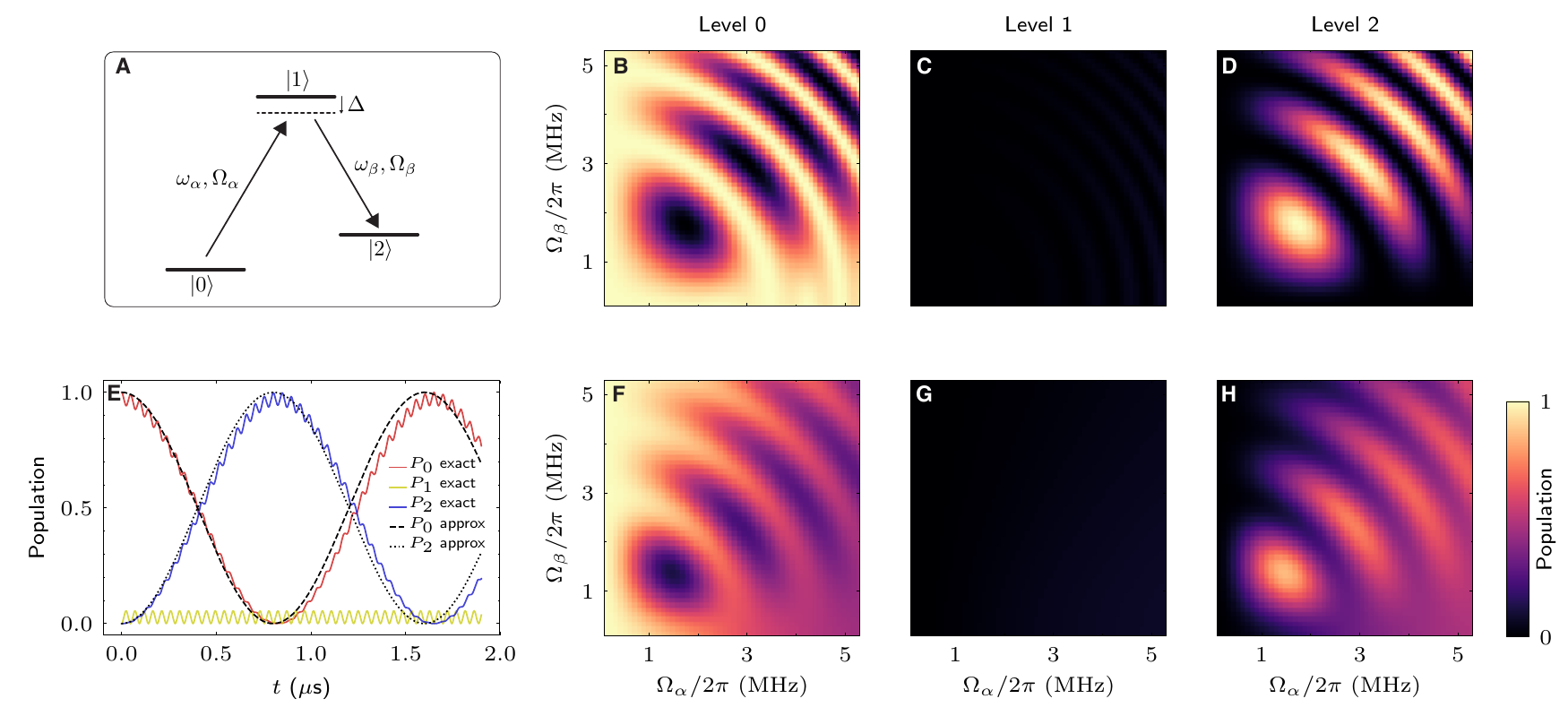}
    \caption{\label{fig:Raman_pulses} (\textbf{A}) Schematic representation of three $0-\pi$ qubit levels coupled to two microwave drives forming a $\Lambda$-system. (\textbf{B} to \textbf{D}) Level population as a function of the drive amplitudes based on the analytical results of the time-evolution of the system ($t=6.7\ \mu$s, $\Delta / 2 \pi$ = 20 MHz). (\textbf{E}) Level population as a function of time based on the exact results for the effective two level system ($\Omega_\alpha / 2 \pi=\Omega_\beta / 2 \pi=$ 5 MHz, $\Delta / 2 \pi$ = 20 MHz). (\textbf{F} to \textbf{H}) Results of the numerical simulation for two Gaussian pulses with $\sigma=1\ \mu$s, $\Delta / 2 \pi$ = 3 MHz, relaxation rates of $\Gamma_{10} / 2 \pi= \Gamma_{12}/ 2 \pi =$ 100 kHz and dephasing rate of $\Gamma_1^\phi / 2 \pi=$ 500 kHz.}
\end{figure}

We model the Raman pulse scheme in the $0-\pi$ qubit by truncating the energy level structure to the ground states of the valleys $|0_s\rangle$, $|\pi_s^+\rangle$ and the intermediate level $|\pi_{d\theta}^-\rangle$. For simplicity, we relabel these levels by $|0_s\rangle\to|0\rangle$, $|\pi_s^+\rangle\to|2\rangle$ and $|\pi_{d\theta}^-\rangle\to|1\rangle$ (see \cref{fig:Raman_pulses}A). We first consider the unitary evolution of this $\Lambda$-system driven by two classical fields (\textit{44})
\begin{equation}
    H/\hbar = \omega_1 \sigma_{11} + \omega_2 \sigma_{22} + \left[\Omega_\alpha \cos{\left(\omega_\alpha t\right)} \sigma_{01} + \Omega_\beta\cos{\left(\omega_\beta t\right)} \sigma_{12}  + h.c.\right],
    \label{eq:HLambda}
\end{equation}
where $\omega_0=0<\omega_2<\omega_1$ are the eigenfrequencies of $|0\rangle$, $|2\rangle$ and $|1\rangle$, respectively, $\omega_\alpha$ and $\omega_\beta$ are the frequencies of the drive tones with amplitudes $\Omega_\alpha$ and $\Omega_\beta$, respectively, while $\sigma_{ij} = |i \rangle \langle j|$ for $i,j\in[1,2,3]$. We moreover assume that the $\alpha$ ($\beta$) drive addresses only the $|0\rangle\leftrightarrow |1\rangle$ ($|1\rangle\leftrightarrow|2\rangle$) transition.

Moving to a rotating frame where the drives are equally detuned from the ancillary level $|1\rangle$, i.e. $\omega_\alpha = \omega_1 - \Delta$, $\omega_\beta=\omega_1-\omega_2-\Delta$, and performing the RWA approximation, \cref{eq:HLambda} takes the time-independent form of 
\begin{equation}
    \Tilde{H}/\hbar = \Delta\sigma_{11} + \left[\frac{1}{2}\Omega_\alpha\sigma_{01} + \frac{1}{2}\Omega_\beta\sigma_{12}+h.c.\right].
    \label{eq:HLambdarot}
\end{equation}
Defining $\Tilde\Omega=\sqrt{\Delta^2 + \Omega_\alpha^2 + \Omega_\beta^2}$, the eigenfrequencies of \cref{eq:HLambdarot} are given by 
\begin{equation}
\begin{split}
    \epsilon_0 &= 0,\\
    \epsilon_\pm &= \frac{1}{2}\left(\Delta\pm \Tilde\Omega\right),
\end{split}
\end{equation}
and correspond to the dressed states 
\begin{equation}
\begin{split}
    |\Psi_0\rangle &= -\Omega_\beta|0\rangle + \Omega_\alpha|2\rangle,\\
    |\Psi_\pm\rangle &=\Omega_\alpha|0\rangle+(\Delta\pm\Tilde\Omega)|1\rangle + \Omega_\beta|2\rangle,
\end{split}
\end{equation}
respectively.

We assume that the system at $t =0$ is initialized in the $|0\rangle$ state when the drives are instantaneously turned on (square pulse). The time-evolution of the system can be obtained by using a basis transformation into the dressed basis. The system evolves to the state $|\Psi(t)\rangle = \alpha(t)|0\rangle + \beta(t)|1\rangle + \gamma(t)|2\rangle$ at time $t$, where
\begin{equation}
\begin{split}
    \alpha(t) &= \frac{\Omega_\alpha^2}{\Omega_\alpha^2 + \Omega_\beta^2}\times\left[\frac{\Omega_\beta^2}{\Omega_\alpha^2}+e^{-i\Delta t/2}\left(\cos\frac{\Tilde\Omega t}{2}+i\frac{\Delta}{\Tilde\Omega}\sin\frac{\Tilde\Omega t}{2}\right)\right],\\
    \beta(t) &=\frac{\Omega_\alpha}{\Tilde\Omega}\times\left[-ie^{-i\Delta t/2}\sin\frac{\Tilde\Omega t}{2} \right],\\
    \gamma(t) &= \frac{\Omega_\alpha\Omega_\beta}{\Omega_\alpha^2 + \Omega_\beta^2}\times\left[-1+e^{-i\Delta t/2}\left(\cos\frac{\Tilde\Omega t}{2}+i\frac{\Delta}{\Tilde\Omega}\sin\frac{\Tilde\Omega t}{2}\right)\right].
\end{split}
\end{equation}

\cref{fig:Raman_pulses}E shows the level populations as a function of time for $\Omega_1=\Omega_2$. We observe Rabi oscillations between the two ground states $|0\rangle$ and $|2\rangle$ with only a negligible population in the intermediate level $|1\rangle$. Interestingly, the Rabi oscillation features a superimposed low amplitude, high frequency modulation (\textit{44}). We note that adiabatic elimination of the intermediate level in the vicinity of equal drives $\Omega_1\approx\Omega_2$ leads to an effective two-level system (\textit{43}) with Rabi rate of $\Omega_R=\Omega_1\Omega_2/2\Delta$. This effective model is in good agreement with the exact analytical solution (dashed and dotted lines in \cref{fig:Raman_pulses}E).

\cref{fig:Raman_pulses}B to D show the level population at a given time as function of the drive amplitude and detuning, similar to the pulsed measurements carried out in our experiment. The results show that maximal population transfer between $|0\rangle$ and $|2\rangle$ is possible when the drives are equal. 

Additionally to the exact solutions, we carried out numerical simulations using the QuTiP software package (\textit{47}) to solve the time evolution of the system involving Gaussian-shaped pulses and decay mechanisms using a Lindblad Master-equation solver. The result of the numerical simulation is in very good agreement with our experimental findings, see Fig.~4C and \cref{fig:Raman_pulses}F to H.

\subsection{Coherence times as a function of external flux}

\begin{figure}[h!]
    \centering
    \includegraphics[width=6cm]{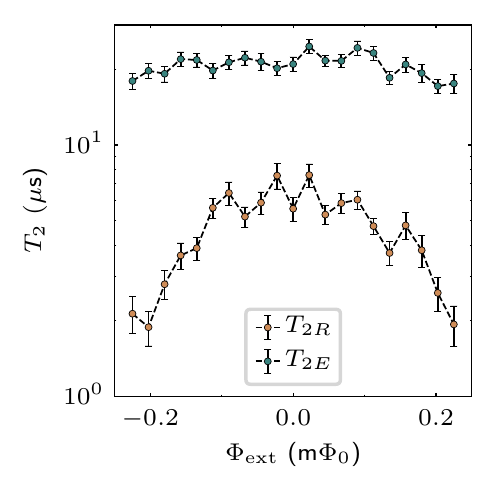}
    \caption{\label{fig:coherence_flux} Measured $T_{2R}$ and $T_{2E}$ values around the magnetic sweet spot. Dashed lines are a guide to the eye.}
\end{figure}

We mapped out the flux-dependence of the coherence times of the logical qubit states in the close vicinity of $\Phi_\mathrm{ext}=0$ (\cref{fig:coherence_flux}). The data demonstrate that the Ramsey coherence times have strong dependence on the magnetic flux with a significant enhancement around the sweet spot. 

\subsection{Autler-Townes spectroscopy as a function of the offset charge}

For completeness, we report all measured Autler-Townes spectroscopy maps obtained at different offset-charge bias, in addition to the one presented in Fig.~3I. As discussed in the main text, we use a strong drive to dress the $|0_{p\theta}\rangle\leftrightarrow|\pi_{p\theta}^-\rangle$ transition. Denoting the qubit transition by $\omega_q$, the coupling rate by $\Omega_c$ and the coupler drive frequency by $\omega_c$, the dispersion of the dressed states takes the form of $\epsilon_\pm=(\omega_q - \omega_c)\pm\sqrt{(\omega_q - \omega_c)^2 + \Omega_c^2}$, which can be measured by an additional weak probe tone. 
\begin{figure}[h!]
    \centering
    \includegraphics[width=17cm]{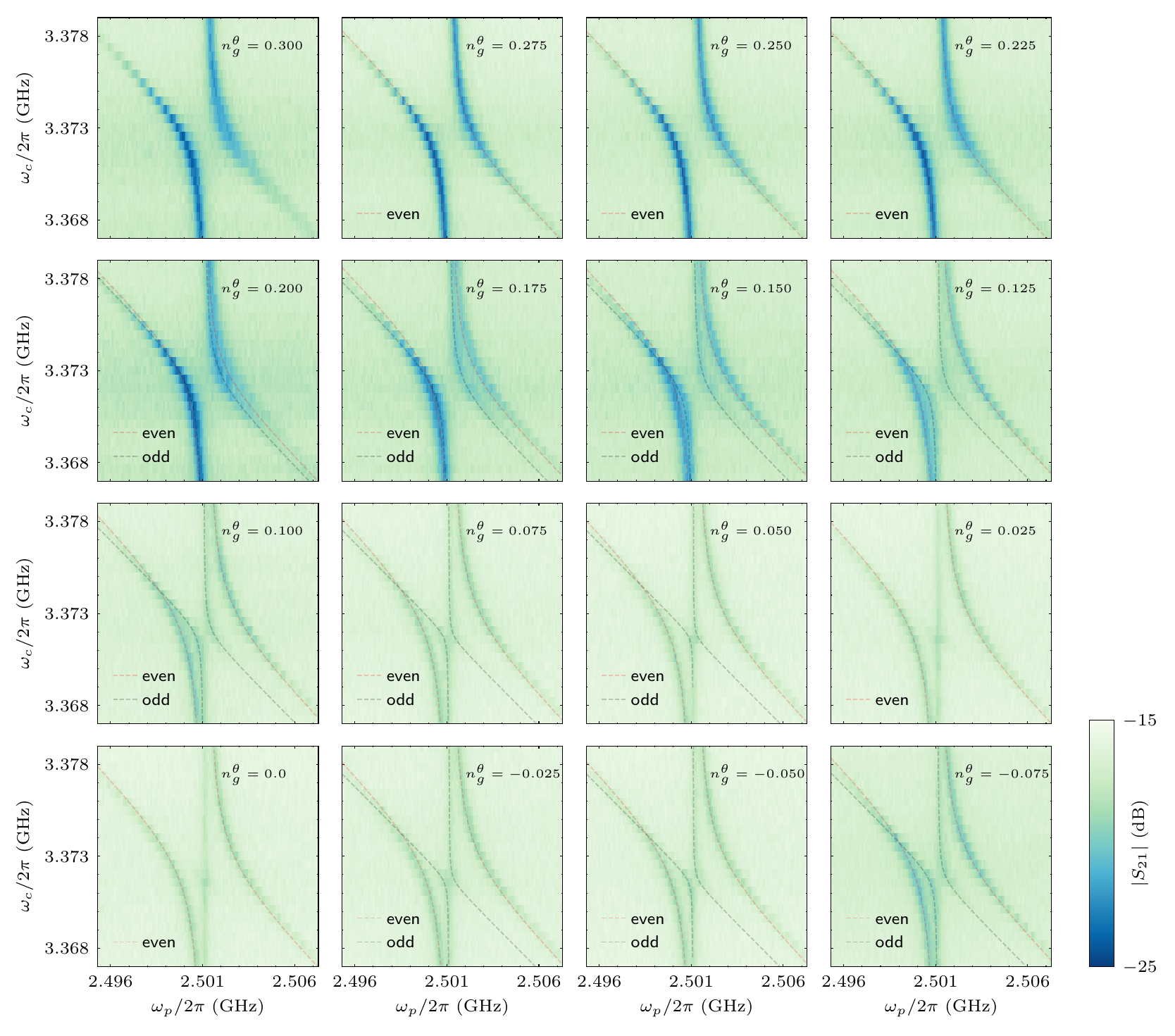}
    \caption{\label{fig:AT_scans} Raw data of the Autler-Townes spectroscopy as a function of the offset-charge. Black and red dashed lines show the least-squares fit to the data in case of odd and even charge parity.}
\end{figure}


